\documentclass[aps,prl,twocolumn,superscriptaddress,a4paper,nopacs]{revtex4}

\usepackage{amsmath}
\usepackage{units}

\usepackage[colorlinks=true,citecolor=blue,linkcolor=magenta]{hyperref}
\hypersetup{
pdfauthor = {},
colorlinks = true, linkcolor = blue, urlcolor=blue, bookmarksnumbered =  true}

\usepackage{graphicx}

\pdfoutput=1 

\begin{document}
\title{A low-loss, broadband antenna for efficient photon collection from a coherent spin in diamond}
\author{D. Riedel}
\altaffiliation{These authors contributed equally to this work}
\affiliation{Department of Physics, University of Basel, Klingelbergstrasse 82, Basel CH-4056, Switzerland}
\author{D. Rohner}
\altaffiliation{These authors contributed equally to this work}
\affiliation{Department of Physics, University of Basel, Klingelbergstrasse 82, Basel CH-4056, Switzerland}
\author{M. Ganzhorn}
\affiliation{Department of Physics, University of Basel, Klingelbergstrasse 82, Basel CH-4056, Switzerland}
\author{T. Kaldewey}
\affiliation{Department of Physics, University of Basel, Klingelbergstrasse 82, Basel CH-4056, Switzerland}
\author{P. Appel}
\affiliation{Department of Physics, University of Basel, Klingelbergstrasse 82, Basel CH-4056, Switzerland}
\author{E. Neu}
\affiliation{Department of Physics, University of Basel, Klingelbergstrasse 82, Basel CH-4056, Switzerland}
\author{R. J. Warburton}
\affiliation{Department of Physics, University of Basel, Klingelbergstrasse 82, Basel CH-4056, Switzerland}
\author{P. Maletinsky}
\email{patrick.maletinsky@unibas.ch}
\affiliation{Department of Physics, University of Basel, Klingelbergstrasse 82, Basel CH-4056, Switzerland}
\date{\today}

\begin{abstract} 

We report the creation of a low-loss, broadband optical antenna giving highly directed output from a coherent single spin in the solid-state. The device, the first solid-state realization of a dielectric antenna, is engineered for individual nitrogen vacancy (NV) electronic spins in diamond. We demonstrate a directionality close to 10. The photonic structure preserves the high spin coherence of single crystal diamond ($T_2\gtrsim100\,\mu$s). The single photon count rate approaches a MHz facilitating efficient spin readout. We thus demonstrate a key enabling technology for quantum applications such as high-sensitivity magnetometry and long-distance spin entanglement. 

\end{abstract}

\maketitle

The electronic spin associated with the nitrogen vacancy (NV) center in diamond constitutes a versatile quantum system with applications in nanoscale magnetometry~\cite{Maze2008,Balasubramanian2008,Maletinsky2012}, quantum communication~\cite{Kurtsiefer2000,Brouri2000,Beveratos2002} and quantum information processing~\cite{Bernien2013}.
NV spins in clean, single-crystalline diamond exhibit remarkable coherence times up to milliseconds even at room temperature~\cite{Balasubramanian2009} and can be initialized and read out optically~\cite{Gruber1997}.
For the majority of applications of NV center spins, the efficiency of collection and detection of the broadband NV photoluminescence (PL) is an essential figure of merit. For example, increased NV PL detection rates lead to improved sensitivities in magnetometry applications~\cite{Rondin2014} and higher two-photon interference rates for entangling remote NV spins~\cite{Bernien2013}.
These collection efficiencies, however, are intrinsically limited by the non-directional emission of NV PL and total internal reflection between the high-index diamond host material and its low-index surrounding. 

To overcome these obstacles and to improve the photon collection efficiency from NV spins, several approaches are currently being pursued. 
Total internal reflection can be reduced by employing solid immersion lenses (SIL)~\cite{Siyushev2010,Marseglia2011} or diamond nanocrystals~\cite{Wolters2012}. 
Conversely, the directionality of NV emission can be improved by optical waveguides~\cite{Babinec2010, Neu2014}, resonators~\cite{Hausmann2012,Faraon2011} or metallic optical antennas~\cite{Choy2011,Wolters2012}. 
However, all these approaches suffer from severe drawbacks. SILs and nanocrystals hardly address the directionality of the NV emission, and diamond nanocrystals generally exhibit poor spin coherence times~\cite{Rondin2014}. Waveguides and resonators only operate in a narrow wavelength range and are therefore ill-suited for broadband emitters such as the NV center and metal-based approaches~\cite{Choy2011,Wolters2012} are plagued by high optical losses. 
Recently, a new, alternative approach based on a layered, dielectric optical antenna was proposed and demonstrated for individual molecules in a low-index polymer matrix~\cite{Lee2011}. Such a ``dielectric optical antenna'' stands out due to its broadband and almost loss-less operation, which can in principle yield near-unity collection efficiencies for single emitters~\cite{Chu2014a}. However, despite their attractiveness, dielectric optical antennas have never been realized for the NV center in diamond where they are particularly well-suited: in addition to their excellent performance, they can be manufactured from high quality single crystal material where the spin coherence is high. The concept is general and versatile: it is potentially powerful for other broadband solid-state emitters.

\begin{figure}
\includegraphics[width=8.6cm]{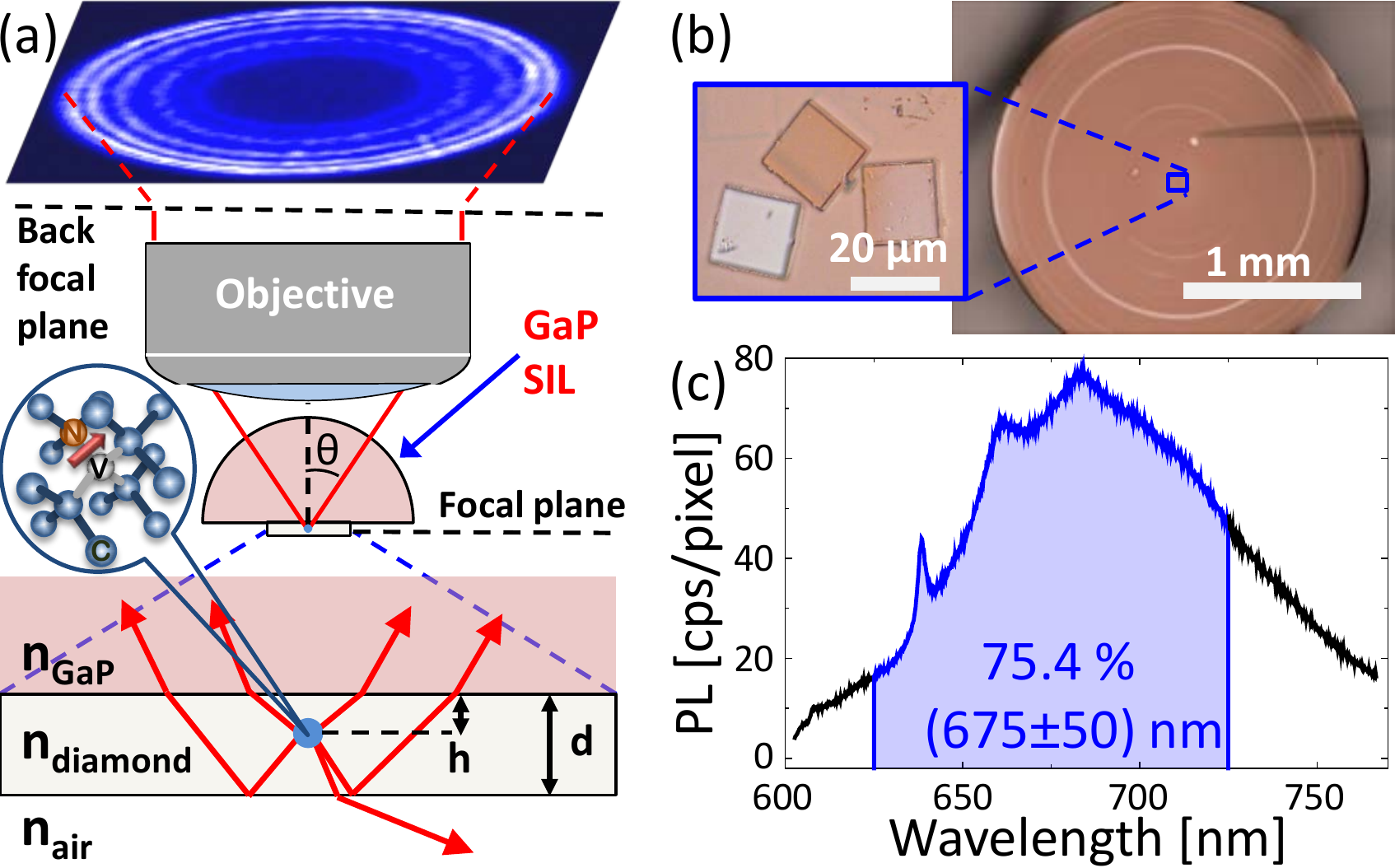}
\caption{\label{schematics} (a) Schematic of the high-index dielectric antenna structure: a thin diamond membrane containing nitrogen vacancy (NV) centers is bonded to a gallium phosphide solid immersion lens (GaP SIL). The majority of the NV photoluminescence (PL) is emitted within the critical angle $\theta \sim 47^\circ$ of the GaP-diamond interface (corresponding to emission within an $\rm{NA}\sim 0.73$).
(b) Positioning of the transferred diamond membranes onto the GaP SIL using a micromanipulator.
(c) Room temperature PL spectrum of a single NV center collected through the dielectric antenna. PL studied in this work passes through a $(675\pm50)~$nm bandpass filter to improve the signal-to-background ratio.}
\end{figure}

In this Letter, we present the first experimental realization of a dielectric optical antenna for an NV center in diamond demonstrating low-loss, broadband operation with a highly directional output, all while preserving the spin coherence of the starting material.  As a central novelty, our work promotes the concept of a dielectric optical antenna for quantum emitters in the solid-state and moreover demonstrates successful antenna-operation for a high-index host material, where light-extraction is particularly challenging. 
Our antenna is based on a thin, single-crystalline diamond membrane directly bonded to the center of a high-index SIL, as illustrated in Fig.\,\ref{schematics}(a). The three layers consisting of air (refractive index $n=1.0$), the diamond film ($n=2.4$) and the gallium phosphide (GaP, $n=3.3$) SIL together form the antenna and lead to the highly directional NV emission.
Crucial to the operation of our antenna is the use of a SIL material with a refractive index higher than that of diamond. This index contrast and the thin diamond membrane are at the heart of the operation of our antenna as they lead to preferential photon emission into the high-index material~\cite{Luan2006}.
The only readily available SIL material with refractive index significantly larger than 2.4, showing transparency over the NV emission spectrum, is GaP\,\cite{Fu2008}, with an index 3.3. The planar interfaces air-diamond-GaP form the dielectric antenna. Efficient out-coupling of the photons from the GaP can be achieved with a hemispherical shape \cite{Wu1999}: photons originating from the antenna have normal incidence at the GaP-air interface, preventing total internal reflection, Fig.\,\ref{schematics}(a). We therefore employ a hemispherical GaP SIL.

In order to obtain a diamond film of suitable thickness for our antenna, we fabricate diamond membranes (with typical dimensions: $\approx 20~\mu$m$~\times20~\mu$m$~\times4~\mu$m) from commercially available single crystal diamond~\cite{Suppl}. Exploiting recently developed top-down diamond nanofabrication techniques~\mbox{\cite{Maletinsky2012, Neu2014}} we structure arrays of free standing membranes with predetermined breaking points~\cite{Suppl}. We then detach the membranes using the sharp tip of a micromanipulator and transfer them to the flat surface of the GaP. The GaP SIL has radius 1.00 mm and was fabricated from bulk material by mechanical polishing.
The correctly positioned diamond membranes bond to the SIL surface by van der Waals forces, Fig.\,\ref{schematics}(b). 
As a last step, we adjust the thicknesses $d$ of the diamond layer by further thinning of the bonded membranes by successive reactive ion etching steps~\cite{Suppl}. 

\begin{figure}
\includegraphics[width=8.6cm]{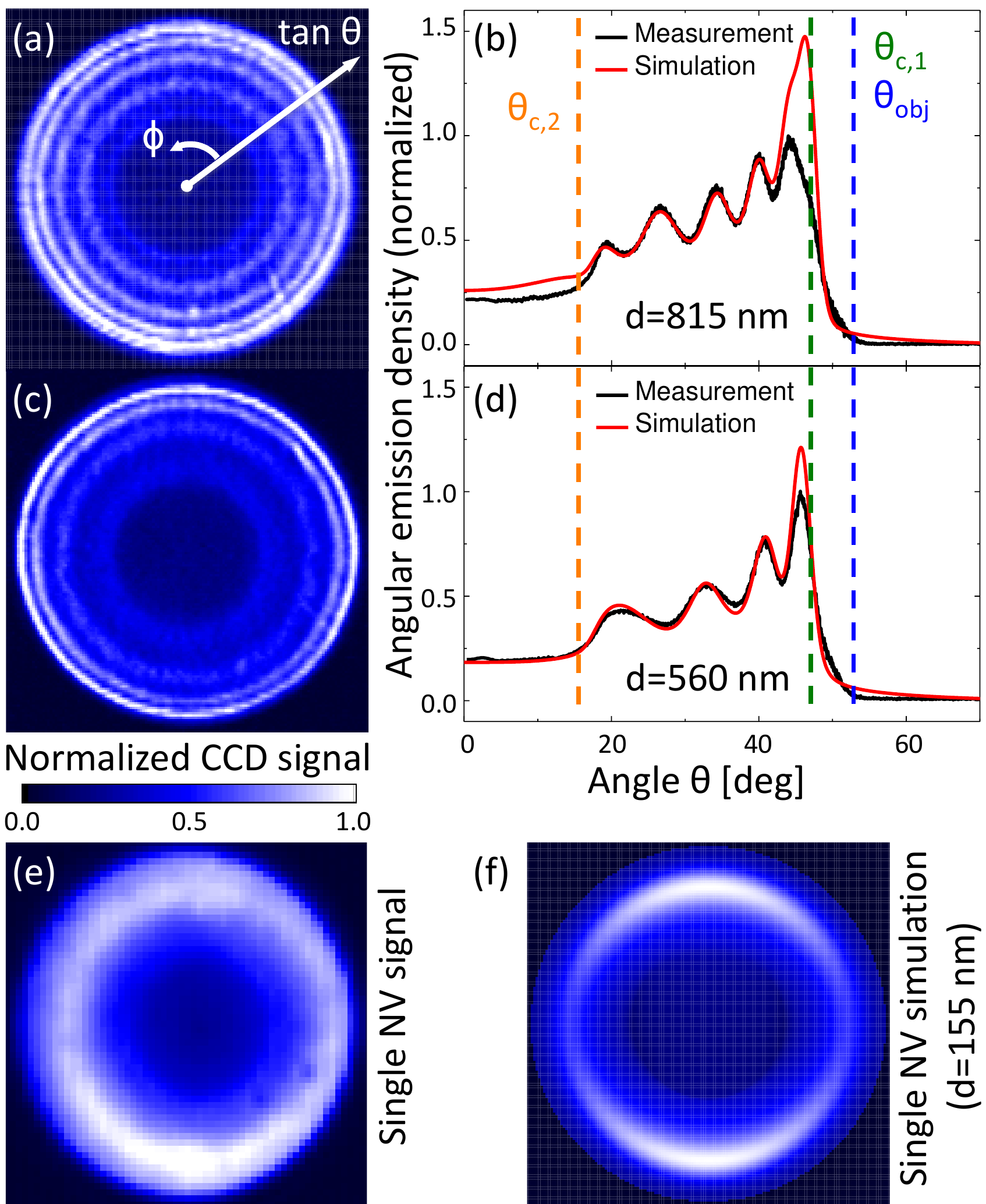}
\caption{\label{BFP} Angular radiation patterns of NV ensembles in the dielectric antenna obtained by back focal plane (BFP) imaging for (a) $d=815~$nm and (c) $d=560~$nm. (b),(d) Average of the measured BFP images over azimuthal angle $\phi$ and calculated emission patterns. (e) Broadband emission pattern of a single NV center. (f) Calculated BFP image for an NV center in a diamond membrane with $d=155~$nm and $h=140~$nm. The color-bar applies to all images in the figure. (a) - (d) use a $(680\pm5)~$nm bandpass filter, (e) a $(675\pm50)~$nm bandpass filter.}
\end{figure}

The PL emission from optical dipoles embedded in dielectric antennas exhibits characteristic radiation patterns~\cite{Lee2011}, signatures of the antenna effects we wish to demonstrate and exploit. We record these radiation patterns for native NV centers in our antenna structures through back focal plane (BFP) imaging. The NV PL is excited using a pulsed, tunable laser source (NKT EXW-12, $\lambda=560~$nm, repetition rate 78 MHz) and detected using a narrowband filter to isolate the strongly wavelength-dependent interference features.
As a first step, we image the PL of NV ensembles in antennas with $d=815~$nm and $d=560~$nm, Fig.\,\ref{BFP}(a),(c).
We observe a radiation pattern consisting of multiple interference rings whose positions strongly vary with $d$. 
For all our BFP images, the NV PL is confined to within a maximal emission angle $\theta_{c,1}$ which corresponds closely to the critical angle at the diamond-GaP interface ($\sim47^\circ$). This observation demonstrates highly directional NV emission into a numerical aperture $\rm{NA}\sim 0.73$. Importantly, this value is significantly smaller than the NA of 0.8 of the microscope objective we employ.
In addition, we observe a region of relatively low PL intensity, bounded by an angle $\theta_{c,2}$, in the center of each BFP image. This region corresponds to PL light escaping the antenna through the diamond-air interface. Due to the collection through the GaP material, the observed value thus corresponds to the critical angle between GaP and air, $\theta_{c,2}\sim18^\circ$. 

For a detailed analysis of the radiation patterns, we average the measured BFP images over the azimuthal angle $\phi$ and compare the resulting emission profile to an analytical calculation, Fig.\,\ref{BFP}(b),(d).
These calculations rely on established procedures for the analysis of the far-field emission from layered, dielectric structures~\cite{Courtois1996, Lieb2004, Luan2006} (for details see Ref.~\cite{Suppl}). 
We use tabulated values for refractive indices of diamond and GaP~\cite{Bass2009} and account for the NV PL spectrum by introducing wavelength-dependent weighting factors for the emission profile which we discretize in steps of $5~$nm.
For both values of $d$, we detect PL from a large ensemble of NVs in the diamond membrane and we assume these NVs to be homogeneously distributed and randomly oriented.
With the exception of the highest emission angles $\theta\sim\theta_{\rm obj}$, our experimental results show excellent agreement with the theoretical expectations.
In particular, the characteristic oscillations we observe allow us to determine precisely the only free parameter in our fit, the thickness $d$ of the diamond layer, which we find to be in good agreement with independent measurements~\cite{Suppl}.

\begin{figure}
\includegraphics[width=8.6cm]{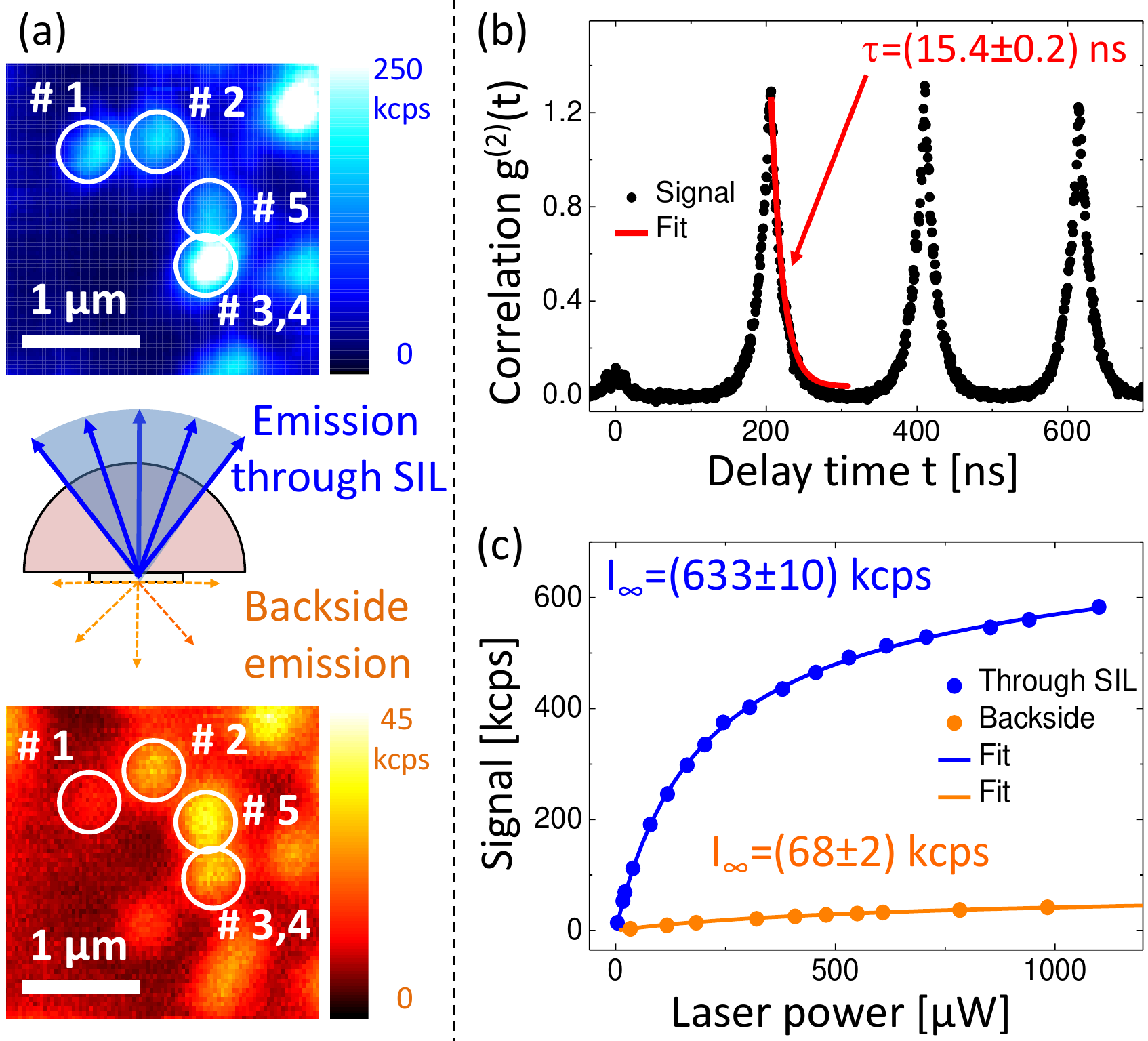}
\caption{\label{singleNV} (a) PL images of individual NV centers recorded through the SIL and through the backside of the antenna. (b) Photon autocorrelation measurement $g^{(2)}(t)$ of NV center \#1 using pulsed excitation. The data shows $g^{(2)}(0)\lesssim 0.1$ and an exponential PL decay with a lifetime $\tau=(15.4\pm0.2)~$ns. (c) Comparison of PL saturation of a single NV in the dielectric antenna when detecting through the GaP SIL and through the backside of the antenna. NV PL is detected using a $(675\pm50)~$nm bandpass filter.}
\end{figure}

Upon further thinning of the diamond membrane, the areal density of NV centers gradually decreases until eventually for a thickness of $d\lesssim150~$nm, we are able to observe NV emission from isolated spots, Fig.\,\ref{singleNV}(a).
Figure\,\ref{BFP}(e) displays the angular radiation pattern of the broadband PL of such a spot.
This emission pattern agrees well with the calculated BFP image for a single NV center, Fig.\,\ref{BFP}(f), which we obtain using only two free parameters: $d$ and the distance $h$ of the NV center from the diamond-GaP interface\,\footnote{We model the single NV by two orthogonal dipoles lying in the plane normal to the NV axis~\cite{Alegre2007}}.
We achieve best agreement between simulation and experiment for $d=155~$nm and $h=140~$nm.

To substantiate further that our observed isolated emission spots indeed stem from single NVs, we determined the photon autocorrelation function $g^{(2)}(t)$ of the detected PL.
Figure\,\ref{singleNV}(b) shows $g^{(2)}(t)$ measured on NV \#1. The data exhibits strong photon antibunching with $g^{(2)}(0)\lesssim 0.1$ and a decay of the pulsed PL on a timescale $\tau=(15.4\pm0.2)~$ns. The decay time is a typical NV center fluorescence lifetime: this value therefore shows that our antenna fabrication process did not induce unwanted non-radiative decay channels for the NV.

To study the directionality $\eta$ of single-photon emission from our antenna, we compare PL emission rates detected through the SIL and through the backside of the antenna, upper and lower panel in Fig.\,\ref{singleNV}(a), respectively.
Our confocal scans reveal several single NVs (labelled $\#1-\#5$ in the figure) scattered throughout the diamond membrane.
Strikingly, NVs appearing bright through the antenna show reduced brightness when observed from the backside and vice versa, indicating that different NVs exhibit different degrees of directionality.
We assign this variation to different vertical positions $h$ for the individual NV centers, which results in varying coupling efficiencies to the antenna mode.

For a more quantitative analysis of the antenna directionality, we compare the saturated PL count rate $I_{\infty}$ of the well-coupled NV \#1, measured through the GaP SIL and from the antenna backside.
We obtain $I_{\infty}$ by fitting the measured PL saturation curves, Fig.\,\ref{singleNV}(c), with $I(P)=I_{\infty} \left( 1+P/P_{\rm{sat}} \right) ^{-1}+b \cdot P$~\cite{Kurtsiefer2000}, where $P$ is the laser power, $P_{\rm{sat}}$ the saturation power and $b \cdot P$ accounts for background fluorescence.
For PL detection through the SIL, we find $I_{\infty,\rm{SIL}}=(633\pm10)~$kcps, while we obtain $I_{\infty,\rm{bs}}=(68\pm2)~$kcps for backside detection\,\footnote{We note that this value of $I_{\infty}$ is similar, but slightly smaller, to the value we find for single NVs measured in unstructured bulk diamond in our setup.}. This corresponds to an enhancement of the detection rate by the antenna which we directly relate to the directionality of our antenna, i.e.\ we find $\eta=9.3$.

We note that the directionality is an underestimate of the directionality of the dielectric antenna itself as $\sim45\%$ of the NV emission is lost on out-coupling, $\sim30\%$ by reflection at the GaP-air interface (a consequence of the abrupt change in reflective index) and a further $\sim10\%$ by scattering losses (a consequence of imperfections in the GaP material~\cite{Suppl}). We also note that a possible source of systematic error is the fact that our current realization of our antenna demands slightly different experimental conditions for top- and bottom-side collection. Specifically, top-side collection was performed in a non-confocal imaging mode since the emission pattern of our antenna has poor overlap with the Gaussian mode of the single mode fiber which represents the ``pin-hole" in our microscope. Backside collection, however, was performed with confocal detection via a single mode fiber since the high density of NVs in our antenna did not allow us to isolate single NVs otherwise. However, we expect back-side confocal detection to be very efficient~\cite{Suppl} such that the systematic error is small. We emphasize that $\eta=9.3$ represents the performance of the entire antenna device without correcting for known losses at the GaP-air interface. For the $d$ and $h$ as determined from the radiation pattern for NV \#1, the NA of the objective lens, and including reflection losses at the GaP-air interface, we calculate a directionality of $\eta = 13.9$, slightly higher than the measured value.

\begin{figure}
\includegraphics[width=8.6cm]{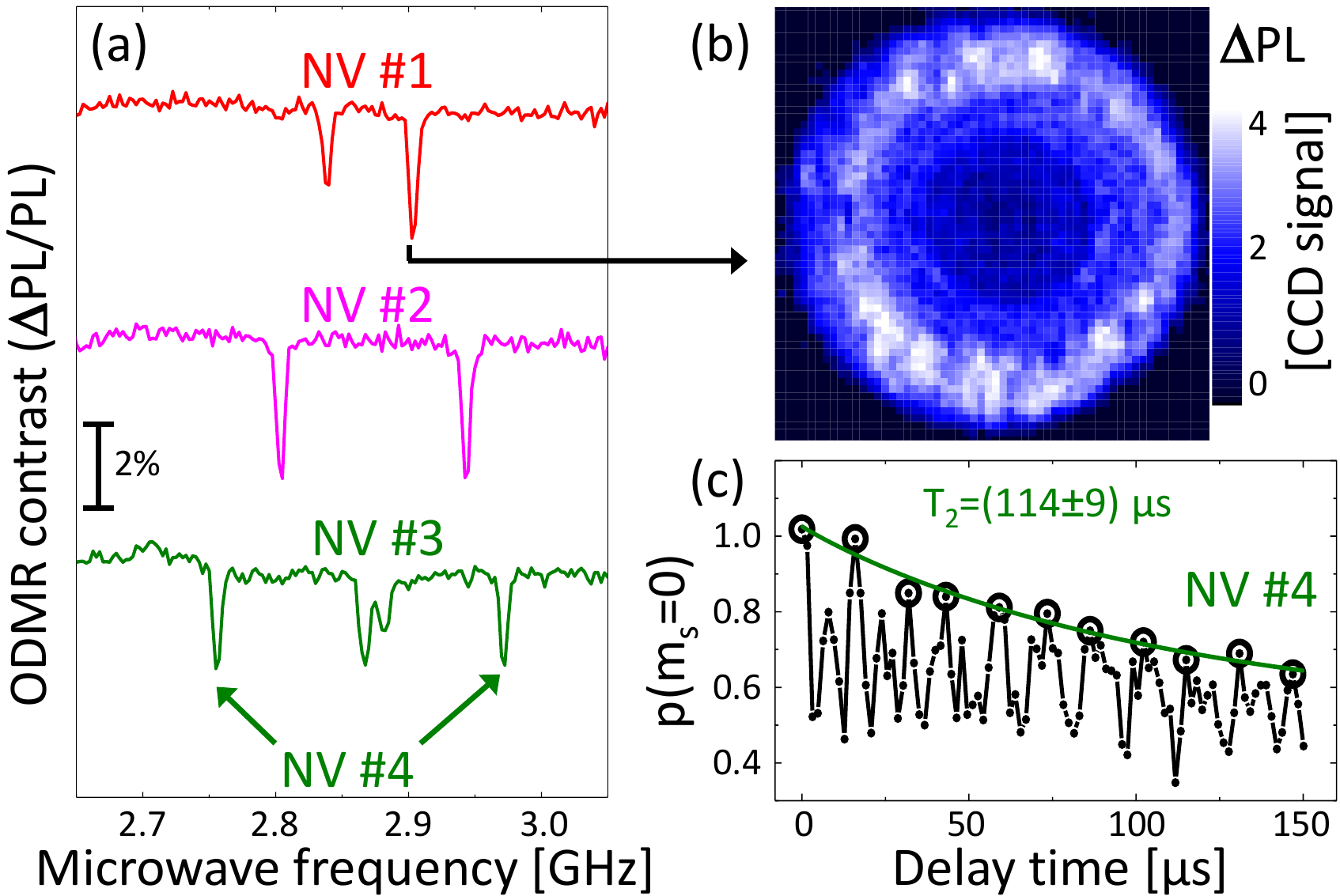}
\caption{\label{ESR} (a) Optically detected magnetic resonance (ODMR) of the NVs highlighted in Fig.\,\ref{singleNV}(a) ($\vert\vec{B}\vert \sim4$~mT), each exhibiting a different orientation within the diamond lattice. The low ODMR contrast results from the low microwave power at the NV location. (b) ODMR amplitude of NV \#1 imaged in the back focal plane. (c) Hahn spin echo measurement on NV \#4 yielding $T_2=(114\pm9)~\mu$s ($\vert\vec{B}\vert \sim11.5$~mT).}
\end{figure}

With single NV centers isolated in our dielectric antenna, we now turn our attention to their electronic spin properties. To that end we perform optically detected magnetic resonance~\cite{Gruber1997} (ODMR) on individual NVs in our antenna. The ODMR resonance frequencies are determined by the strength of an external magnetic field $\vec{B}$ and the orientation between the NV axis and $\vec{B}$. We apply $\vec{B}$ in an oblique direction with respect to the four possible NV axes such that ODMR allows us to differentiate NVs with different orientations. Fig.\,\ref{ESR}(a) shows ODMR for NVs \#1-4 in a static magnetic field. ODMR enables us to probe the spin dependence of the BFP image in Fig.\,\ref{BFP}(c). We record the difference of PL in the BFP when driving the NV on and off spin resonance. The corresponding, spin-resolved BFP image is shown in Fig.\,\ref{ESR}(b) and is, up to a constant scaling factor, identical to Fig.\,\ref{BFP}(c). In particular, the ODMR contrast is constant throughout the BFP and no ODMR contrast is detected in the BFP when we drive the spin resonance of one of the NV \#1's neighbors (all of which have orientations different to NV \#1).

Finally, we use coherent manipulation of the NVs' electronic spins to probe their $T_2$ times (strictly, the Hahn echo coherence times). Fig.\,\ref{ESR}(c) shows a Hahn-echo measurement performed on NV \#4. We find $T_2=(114\pm9)~\mu$s, which is typical for the chemical vapor deposition diamond material we employ here. This demonstrates that fabrication of the dielectric membrane preserves the long NV electronic spin coherence times. 

The radical change in the radiation patterns and the highly directed PL emission from single NVs demonstrate proper operation of the dielectric antenna. However, several factors still limiting its performance could be improved in future experiments. The losses at the GaP-air interface could be avoided primarily by an antireflection coating, but also by smoother surfaces; the losses at large angles need to be investigated and mitigated. A slight improvement can be gained by optimization of $d$ and $h$: according to our calculations, PL collection efficiency can be increased from the present 95\% ($d=155~$nm, $h=140~$nm) to above $99\%$~\cite{Suppl}. Lastly, the emission mode of the antenna can be improved by using (111)-oriented diamond where the NV emission dipoles show an orientation better suited for our antenna~\cite{Neu2014}. We note also that the use of an optimized background filter would readily reduce the losses in our setup which are presently $\sim 25\%$, Fig.\,\ref{schematics}(c). Altogether, an improvement in the NV PL detection efficiency by close to one order of magnitude can be envisaged. 

In summary, we have for the first time applied the powerful concept of a dielectric optical antenna~\cite{Lee2011} to single quantum-emitters in the solid-state.  We have demonstrated successful antenna operation and the addressing of single NV spins in the antenna. The antenna has broadband operation and it preserves the advantageous optical and spin properties of the single crystal diamond starting material. 
With the high antenna-directionality, photon collection efficiency and spin coherence times we were able to demonstrate simultaneously, our results present significant advances over previous approaches in the field of diamond nano-photonics\,\cite{Siyushev2010,Marseglia2011,Wolters2012,Babinec2010, Neu2014,Hausmann2012,Faraon2011,Choy2011}. Further improvements to our structure should allow for record-high single NV count rates in the range of several MHz. 
Our antenna design is immediately applicable to a variety of applications ranging from wide-field magnetic imaging~\cite{Lesage2013} to nanoscale nuclear spin detection~\cite{Staudacher2013} and entanglement of remote spins~\cite{Bernien2013}, where in all cases it is poised to lead to a significant leap in performance. Finally, we emphasize that our approach is not limited to NV centers -- other color centers in diamond~\cite{Neu2011} can also profit from the antenna -- or even to diamond itself: SiC~\cite{Riedel2012a} and other solid-state hosts can be integrated into planar dielectric antennas. 

We thank M.\ Kasperczik and L.\ Novotny for help with our back focal plane imaging experiments. We gratefully acknowledge financial support from SNI; NCCR QSIT; EU FP7 ITN S$^3$NANO; SNF grants 200021\_143697 and 200021\_132313; and EU FP7 grant 611143 (DIADEMS). 

\bibliographystyle{apsrev4-1}

\begin{thebibliography}{32}%
\makeatletter
\providecommand \@ifxundefined [1]{%
 \@ifx{#1\undefined}
}%
\providecommand \@ifnum [1]{%
 \ifnum #1\expandafter \@firstoftwo
 \else \expandafter \@secondoftwo
 \fi
}%
\providecommand \@ifx [1]{%
 \ifx #1\expandafter \@firstoftwo
 \else \expandafter \@secondoftwo
 \fi
}%
\providecommand \natexlab [1]{#1}%
\providecommand \enquote  [1]{``#1''}%
\providecommand \bibnamefont  [1]{#1}%
\providecommand \bibfnamefont [1]{#1}%
\providecommand \citenamefont [1]{#1}%
\providecommand \href@noop [0]{\@secondoftwo}%
\providecommand \href [0]{\begingroup \@sanitize@url \@href}%
\providecommand \@href[1]{\@@startlink{#1}\@@href}%
\providecommand \@@href[1]{\endgroup#1\@@endlink}%
\providecommand \@sanitize@url [0]{\catcode `\\12\catcode `\$12\catcode
  `\&12\catcode `\#12\catcode `\^12\catcode `\_12\catcode `\%12\relax}%
\providecommand \@@startlink[1]{}%
\providecommand \@@endlink[0]{}%
\providecommand \url  [0]{\begingroup\@sanitize@url \@url }%
\providecommand \@url [1]{\endgroup\@href {#1}{\urlprefix }}%
\providecommand \urlprefix  [0]{URL }%
\providecommand \Eprint [0]{\href }%
\providecommand \doibase [0]{http://dx.doi.org/}%
\providecommand \selectlanguage [0]{\@gobble}%
\providecommand \bibinfo  [0]{\@secondoftwo}%
\providecommand \bibfield  [0]{\@secondoftwo}%
\providecommand \translation [1]{[#1]}%
\providecommand \BibitemOpen [0]{}%
\providecommand \bibitemStop [0]{}%
\providecommand \bibitemNoStop [0]{.\EOS\space}%
\providecommand \EOS [0]{\spacefactor3000\relax}%
\providecommand \BibitemShut  [1]{\csname bibitem#1\endcsname}%
\let\auto@bib@innerbib\@empty
\bibitem [{\citenamefont {Maze}\ \emph {et~al.}(2008)\citenamefont {Maze},
  \citenamefont {Stanwix}, \citenamefont {Hodges}, \citenamefont {Hong},
  \citenamefont {Taylor}, \citenamefont {Cappellaro}, \citenamefont {Jiang},
  \citenamefont {Dutt}, \citenamefont {Togan}, \citenamefont {Zibrov},
  \citenamefont {Yacoby}, \citenamefont {Walsworth},\ and\ \citenamefont
  {Lukin}}]{Maze2008}%
  \BibitemOpen
  \bibfield  {author} {\bibinfo {author} {\bibfnamefont {J.~R.}\ \bibnamefont
  {Maze}}, \bibinfo {author} {\bibfnamefont {P.~L.}\ \bibnamefont {Stanwix}},
  \bibinfo {author} {\bibfnamefont {J.~S.}\ \bibnamefont {Hodges}}, \bibinfo
  {author} {\bibfnamefont {S.}~\bibnamefont {Hong}}, \bibinfo {author}
  {\bibfnamefont {J.~M.}\ \bibnamefont {Taylor}}, \bibinfo {author}
  {\bibfnamefont {P.}~\bibnamefont {Cappellaro}}, \bibinfo {author}
  {\bibfnamefont {L.}~\bibnamefont {Jiang}}, \bibinfo {author} {\bibfnamefont
  {M.~V.~G.}\ \bibnamefont {Dutt}}, \bibinfo {author} {\bibfnamefont
  {E.}~\bibnamefont {Togan}}, \bibinfo {author} {\bibfnamefont {A.~S.}\
  \bibnamefont {Zibrov}}, \bibinfo {author} {\bibfnamefont {A.}~\bibnamefont
  {Yacoby}}, \bibinfo {author} {\bibfnamefont {R.~L.}\ \bibnamefont
  {Walsworth}}, \ and\ \bibinfo {author} {\bibfnamefont {M.~D.}\ \bibnamefont
  {Lukin}},\ }\href {\doibase 10.1038/nature07279} {\bibfield  {journal}
  {\bibinfo  {journal} {Nature}\ }\textbf {\bibinfo {volume} {455}},\ \bibinfo
  {pages} {644} (\bibinfo {year} {2008})}\BibitemShut {NoStop}%
\bibitem [{\citenamefont {Balasubramanian}\ \emph {et~al.}(2008)\citenamefont
  {Balasubramanian}, \citenamefont {Chan}, \citenamefont {Kolesov},
  \citenamefont {Al-Hmoud}, \citenamefont {Tisler}, \citenamefont {Shin},
  \citenamefont {Kim}, \citenamefont {Wojcik}, \citenamefont {Hemmer},
  \citenamefont {Krueger}, \citenamefont {Hanke}, \citenamefont
  {Leitenstorfer}, \citenamefont {Bratschitsch}, \citenamefont {Jelezko},\ and\
  \citenamefont {Wrachtrup}}]{Balasubramanian2008}%
  \BibitemOpen
  \bibfield  {author} {\bibinfo {author} {\bibfnamefont {G.}~\bibnamefont
  {Balasubramanian}}, \bibinfo {author} {\bibfnamefont {I.~Y.}\ \bibnamefont
  {Chan}}, \bibinfo {author} {\bibfnamefont {R.}~\bibnamefont {Kolesov}},
  \bibinfo {author} {\bibfnamefont {M.}~\bibnamefont {Al-Hmoud}}, \bibinfo
  {author} {\bibfnamefont {J.}~\bibnamefont {Tisler}}, \bibinfo {author}
  {\bibfnamefont {C.}~\bibnamefont {Shin}}, \bibinfo {author} {\bibfnamefont
  {C.}~\bibnamefont {Kim}}, \bibinfo {author} {\bibfnamefont {A.}~\bibnamefont
  {Wojcik}}, \bibinfo {author} {\bibfnamefont {P.~R.}\ \bibnamefont {Hemmer}},
  \bibinfo {author} {\bibfnamefont {A.}~\bibnamefont {Krueger}}, \bibinfo
  {author} {\bibfnamefont {T.}~\bibnamefont {Hanke}}, \bibinfo {author}
  {\bibfnamefont {A.}~\bibnamefont {Leitenstorfer}}, \bibinfo {author}
  {\bibfnamefont {R.}~\bibnamefont {Bratschitsch}}, \bibinfo {author}
  {\bibfnamefont {F.}~\bibnamefont {Jelezko}}, \ and\ \bibinfo {author}
  {\bibfnamefont {J.}~\bibnamefont {Wrachtrup}},\ }\href {\doibase
  10.1038/nature07278} {\bibfield  {journal} {\bibinfo  {journal} {Nature}\
  }\textbf {\bibinfo {volume} {455}},\ \bibinfo {pages} {648} (\bibinfo {year}
  {2008})}\BibitemShut {NoStop}%
\bibitem [{\citenamefont {Maletinsky}\ \emph {et~al.}(2012)\citenamefont
  {Maletinsky}, \citenamefont {Hong}, \citenamefont {Grinolds}, \citenamefont
  {Hausmann}, \citenamefont {Lukin}, \citenamefont {Walsworth}, \citenamefont
  {Loncar},\ and\ \citenamefont {Yacoby}}]{Maletinsky2012}%
  \BibitemOpen
  \bibfield  {author} {\bibinfo {author} {\bibfnamefont {P.}~\bibnamefont
  {Maletinsky}}, \bibinfo {author} {\bibfnamefont {S.}~\bibnamefont {Hong}},
  \bibinfo {author} {\bibfnamefont {M.~S.}\ \bibnamefont {Grinolds}}, \bibinfo
  {author} {\bibfnamefont {B.}~\bibnamefont {Hausmann}}, \bibinfo {author}
  {\bibfnamefont {M.~D.}\ \bibnamefont {Lukin}}, \bibinfo {author}
  {\bibfnamefont {R.~L.}\ \bibnamefont {Walsworth}}, \bibinfo {author}
  {\bibfnamefont {M.}~\bibnamefont {Loncar}}, \ and\ \bibinfo {author}
  {\bibfnamefont {A.}~\bibnamefont {Yacoby}},\ }\href {\doibase
  10.1038/nnano.2012.50} {\bibfield  {journal} {\bibinfo  {journal} {Nat.
  Nanotechnol.}\ }\textbf {\bibinfo {volume} {7}},\ \bibinfo {pages} {320}
  (\bibinfo {year} {2012})}\BibitemShut {NoStop}%
\bibitem [{\citenamefont {Kurtsiefer}\ \emph {et~al.}(2000)\citenamefont
  {Kurtsiefer}, \citenamefont {Mayer}, \citenamefont {Zarda},\ and\
  \citenamefont {Weinfurter}}]{Kurtsiefer2000}%
  \BibitemOpen
  \bibfield  {author} {\bibinfo {author} {\bibfnamefont {C.}~\bibnamefont
  {Kurtsiefer}}, \bibinfo {author} {\bibfnamefont {S.}~\bibnamefont {Mayer}},
  \bibinfo {author} {\bibfnamefont {P.}~\bibnamefont {Zarda}}, \ and\ \bibinfo
  {author} {\bibfnamefont {H.}~\bibnamefont {Weinfurter}},\ }\href {\doibase
  10.1103/PhysRevLett.85.290} {\bibfield  {journal} {\bibinfo  {journal} {Phys.
  Rev. Lett.}\ }\textbf {\bibinfo {volume} {85}},\ \bibinfo {pages} {290}
  (\bibinfo {year} {2000})}\BibitemShut {NoStop}%
\bibitem [{\citenamefont {Brouri}\ \emph {et~al.}(2000)\citenamefont {Brouri},
  \citenamefont {Beveratos}, \citenamefont {Poizat},\ and\ \citenamefont
  {Grangier}}]{Brouri2000}%
  \BibitemOpen
  \bibfield  {author} {\bibinfo {author} {\bibfnamefont {R.}~\bibnamefont
  {Brouri}}, \bibinfo {author} {\bibfnamefont {A.}~\bibnamefont {Beveratos}},
  \bibinfo {author} {\bibfnamefont {J.~P.}\ \bibnamefont {Poizat}}, \ and\
  \bibinfo {author} {\bibfnamefont {P.}~\bibnamefont {Grangier}},\ }\href
  {http://www.ncbi.nlm.nih.gov/pubmed/18066197} {\bibfield  {journal} {\bibinfo
   {journal} {Opt. Lett.}\ }\textbf {\bibinfo {volume} {25}},\ \bibinfo {pages}
  {1294} (\bibinfo {year} {2000})}\BibitemShut {NoStop}%
\bibitem [{\citenamefont {Beveratos}\ \emph {et~al.}(2002)\citenamefont
  {Beveratos}, \citenamefont {Brouri}, \citenamefont {Gacoin}, \citenamefont
  {Villing}, \citenamefont {Poizat},\ and\ \citenamefont
  {Grangier}}]{Beveratos2002}%
  \BibitemOpen
  \bibfield  {author} {\bibinfo {author} {\bibfnamefont {A.}~\bibnamefont
  {Beveratos}}, \bibinfo {author} {\bibfnamefont {R.}~\bibnamefont {Brouri}},
  \bibinfo {author} {\bibfnamefont {T.}~\bibnamefont {Gacoin}}, \bibinfo
  {author} {\bibfnamefont {A.}~\bibnamefont {Villing}}, \bibinfo {author}
  {\bibfnamefont {J.-P.}\ \bibnamefont {Poizat}}, \ and\ \bibinfo {author}
  {\bibfnamefont {P.}~\bibnamefont {Grangier}},\ }\href {\doibase
  10.1103/PhysRevLett.89.187901} {\bibfield  {journal} {\bibinfo  {journal}
  {Phys. Rev. Lett.}\ }\textbf {\bibinfo {volume} {89}},\ \bibinfo {pages}
  {187901} (\bibinfo {year} {2002})}\BibitemShut {NoStop}%
\bibitem [{\citenamefont {Bernien}\ \emph {et~al.}(2013)\citenamefont
  {Bernien}, \citenamefont {Hensen}, \citenamefont {Pfaff}, \citenamefont
  {Koolstra}, \citenamefont {Blok}, \citenamefont {Robledo}, \citenamefont
  {Taminiau}, \citenamefont {Markham}, \citenamefont {Twitchen}, \citenamefont
  {Childress},\ and\ \citenamefont {Hanson}}]{Bernien2013}%
  \BibitemOpen
  \bibfield  {author} {\bibinfo {author} {\bibfnamefont {H.}~\bibnamefont
  {Bernien}}, \bibinfo {author} {\bibfnamefont {B.}~\bibnamefont {Hensen}},
  \bibinfo {author} {\bibfnamefont {W.}~\bibnamefont {Pfaff}}, \bibinfo
  {author} {\bibfnamefont {G.}~\bibnamefont {Koolstra}}, \bibinfo {author}
  {\bibfnamefont {M.~S.}\ \bibnamefont {Blok}}, \bibinfo {author}
  {\bibfnamefont {L.}~\bibnamefont {Robledo}}, \bibinfo {author} {\bibfnamefont
  {T.~H.}\ \bibnamefont {Taminiau}}, \bibinfo {author} {\bibfnamefont
  {M.}~\bibnamefont {Markham}}, \bibinfo {author} {\bibfnamefont {D.~J.}\
  \bibnamefont {Twitchen}}, \bibinfo {author} {\bibfnamefont {L.}~\bibnamefont
  {Childress}}, \ and\ \bibinfo {author} {\bibfnamefont {R.}~\bibnamefont
  {Hanson}},\ }\href {\doibase 10.1038/nature12016} {\bibfield  {journal}
  {\bibinfo  {journal} {Nature}\ }\textbf {\bibinfo {volume} {497}},\ \bibinfo
  {pages} {86} (\bibinfo {year} {2013})}\BibitemShut {NoStop}%
\bibitem [{\citenamefont {Balasubramanian}\ \emph {et~al.}(2009)\citenamefont
  {Balasubramanian}, \citenamefont {Neumann}, \citenamefont {Twitchen},
  \citenamefont {Markham}, \citenamefont {Kolesov}, \citenamefont {Mizuochi},
  \citenamefont {Isoya}, \citenamefont {Achard}, \citenamefont {Beck},
  \citenamefont {Tissler}, \citenamefont {Jacques}, \citenamefont {Hemmer},
  \citenamefont {Jelezko},\ and\ \citenamefont
  {Wrachtrup}}]{Balasubramanian2009}%
  \BibitemOpen
  \bibfield  {author} {\bibinfo {author} {\bibfnamefont {G.}~\bibnamefont
  {Balasubramanian}}, \bibinfo {author} {\bibfnamefont {P.}~\bibnamefont
  {Neumann}}, \bibinfo {author} {\bibfnamefont {D.}~\bibnamefont {Twitchen}},
  \bibinfo {author} {\bibfnamefont {M.}~\bibnamefont {Markham}}, \bibinfo
  {author} {\bibfnamefont {R.}~\bibnamefont {Kolesov}}, \bibinfo {author}
  {\bibfnamefont {N.}~\bibnamefont {Mizuochi}}, \bibinfo {author}
  {\bibfnamefont {J.}~\bibnamefont {Isoya}}, \bibinfo {author} {\bibfnamefont
  {J.}~\bibnamefont {Achard}}, \bibinfo {author} {\bibfnamefont
  {J.}~\bibnamefont {Beck}}, \bibinfo {author} {\bibfnamefont {J.}~\bibnamefont
  {Tissler}}, \bibinfo {author} {\bibfnamefont {V.}~\bibnamefont {Jacques}},
  \bibinfo {author} {\bibfnamefont {P.~R.}\ \bibnamefont {Hemmer}}, \bibinfo
  {author} {\bibfnamefont {F.}~\bibnamefont {Jelezko}}, \ and\ \bibinfo
  {author} {\bibfnamefont {J.}~\bibnamefont {Wrachtrup}},\ }\href {\doibase
  10.1038/nmat2420} {\bibfield  {journal} {\bibinfo  {journal} {Nat. Mater.}\
  }\textbf {\bibinfo {volume} {8}},\ \bibinfo {pages} {383} (\bibinfo {year}
  {2009})}\BibitemShut {NoStop}%
\bibitem [{\citenamefont {Gruber}\ \emph {et~al.}(1997)\citenamefont {Gruber},
  \citenamefont {Dr\"{a}benstedt}, \citenamefont {Tietz}, \citenamefont
  {Fleury}, \citenamefont {Wrachtrup},\ and\ \citenamefont {von
  Borczyskowski}}]{Gruber1997}%
  \BibitemOpen
  \bibfield  {author} {\bibinfo {author} {\bibfnamefont {A.}~\bibnamefont
  {Gruber}}, \bibinfo {author} {\bibfnamefont {A.}~\bibnamefont
  {Dr\"{a}benstedt}}, \bibinfo {author} {\bibfnamefont {C.}~\bibnamefont
  {Tietz}}, \bibinfo {author} {\bibfnamefont {L.}~\bibnamefont {Fleury}},
  \bibinfo {author} {\bibfnamefont {J.}~\bibnamefont {Wrachtrup}}, \ and\
  \bibinfo {author} {\bibfnamefont {C.}~\bibnamefont {von Borczyskowski}},\
  }\href {\doibase 10.1126/science.276.5321.2012} {\bibfield  {journal}
  {\bibinfo  {journal} {Science}\ }\textbf {\bibinfo {volume} {276}},\ \bibinfo
  {pages} {2012} (\bibinfo {year} {1997})}\BibitemShut {NoStop}%
\bibitem [{\citenamefont {Rondin}\ \emph {et~al.}(2014)\citenamefont {Rondin},
  \citenamefont {Tetienne}, \citenamefont {Hingant}, \citenamefont {Roch},
  \citenamefont {Maletinsky},\ and\ \citenamefont {Jacques}}]{Rondin2014}%
  \BibitemOpen
  \bibfield  {author} {\bibinfo {author} {\bibfnamefont {L.}~\bibnamefont
  {Rondin}}, \bibinfo {author} {\bibfnamefont {J.-P.}\ \bibnamefont
  {Tetienne}}, \bibinfo {author} {\bibfnamefont {T.}~\bibnamefont {Hingant}},
  \bibinfo {author} {\bibfnamefont {J.-F.}\ \bibnamefont {Roch}}, \bibinfo
  {author} {\bibfnamefont {P.}~\bibnamefont {Maletinsky}}, \ and\ \bibinfo
  {author} {\bibfnamefont {V.}~\bibnamefont {Jacques}},\ }\href {\doibase
  10.1088/0034-4885/77/5/056503} {\bibfield  {journal} {\bibinfo  {journal}
  {Rep. Prog. Phys.}\ }\textbf {\bibinfo {volume} {77}},\ \bibinfo {pages}
  {056503} (\bibinfo {year} {2014})}\BibitemShut {NoStop}%
\bibitem [{\citenamefont {Siyushev}\ \emph {et~al.}(2010)\citenamefont
  {Siyushev}, \citenamefont {Kaiser}, \citenamefont {Jacques}, \citenamefont
  {Gerhardt}, \citenamefont {Bischof}, \citenamefont {Fedder}, \citenamefont
  {Dodson}, \citenamefont {Markham}, \citenamefont {Twitchen}, \citenamefont
  {Jelezko},\ and\ \citenamefont {Wrachtrup}}]{Siyushev2010}%
  \BibitemOpen
  \bibfield  {author} {\bibinfo {author} {\bibfnamefont {P.}~\bibnamefont
  {Siyushev}}, \bibinfo {author} {\bibfnamefont {F.}~\bibnamefont {Kaiser}},
  \bibinfo {author} {\bibfnamefont {V.}~\bibnamefont {Jacques}}, \bibinfo
  {author} {\bibfnamefont {I.}~\bibnamefont {Gerhardt}}, \bibinfo {author}
  {\bibfnamefont {S.}~\bibnamefont {Bischof}}, \bibinfo {author} {\bibfnamefont
  {H.}~\bibnamefont {Fedder}}, \bibinfo {author} {\bibfnamefont
  {J.}~\bibnamefont {Dodson}}, \bibinfo {author} {\bibfnamefont
  {M.}~\bibnamefont {Markham}}, \bibinfo {author} {\bibfnamefont
  {D.}~\bibnamefont {Twitchen}}, \bibinfo {author} {\bibfnamefont
  {F.}~\bibnamefont {Jelezko}}, \ and\ \bibinfo {author} {\bibfnamefont
  {J.}~\bibnamefont {Wrachtrup}},\ }\href {\doibase 10.1063/1.3519849}
  {\bibfield  {journal} {\bibinfo  {journal} {Appl. Phys. Lett.}\ }\textbf
  {\bibinfo {volume} {97}},\ \bibinfo {pages} {241902} (\bibinfo {year}
  {2010})}\BibitemShut {NoStop}%
\bibitem [{\citenamefont {Marseglia}\ \emph {et~al.}(2011)\citenamefont
  {Marseglia}, \citenamefont {Hadden}, \citenamefont {Stanley-Clarke},
  \citenamefont {Harrison}, \citenamefont {Patton}, \citenamefont {Ho},
  \citenamefont {Naydenov}, \citenamefont {Jelezko}, \citenamefont {Meijer},
  \citenamefont {Dolan}, \citenamefont {Smith}, \citenamefont {Rarity},\ and\
  \citenamefont {O'Brien}}]{Marseglia2011}%
  \BibitemOpen
  \bibfield  {author} {\bibinfo {author} {\bibfnamefont {L.}~\bibnamefont
  {Marseglia}}, \bibinfo {author} {\bibfnamefont {J.~P.}\ \bibnamefont
  {Hadden}}, \bibinfo {author} {\bibfnamefont {A.~C.}\ \bibnamefont
  {Stanley-Clarke}}, \bibinfo {author} {\bibfnamefont {J.~P.}\ \bibnamefont
  {Harrison}}, \bibinfo {author} {\bibfnamefont {B.}~\bibnamefont {Patton}},
  \bibinfo {author} {\bibfnamefont {Y.-L.~D.}\ \bibnamefont {Ho}}, \bibinfo
  {author} {\bibfnamefont {B.}~\bibnamefont {Naydenov}}, \bibinfo {author}
  {\bibfnamefont {F.}~\bibnamefont {Jelezko}}, \bibinfo {author} {\bibfnamefont
  {J.}~\bibnamefont {Meijer}}, \bibinfo {author} {\bibfnamefont {P.~R.}\
  \bibnamefont {Dolan}}, \bibinfo {author} {\bibfnamefont {J.~M.}\ \bibnamefont
  {Smith}}, \bibinfo {author} {\bibfnamefont {J.~G.}\ \bibnamefont {Rarity}}, \
  and\ \bibinfo {author} {\bibfnamefont {J.~L.}\ \bibnamefont {O'Brien}},\
  }\href {\doibase 10.1063/1.3573870} {\bibfield  {journal} {\bibinfo
  {journal} {Appl. Phys. Lett.}\ }\textbf {\bibinfo {volume} {98}},\ \bibinfo
  {pages} {133107} (\bibinfo {year} {2011})}\BibitemShut {NoStop}%
\bibitem [{\citenamefont {Wolters}\ \emph {et~al.}(2012)\citenamefont
  {Wolters}, \citenamefont {Kewes}, \citenamefont {Schell}, \citenamefont
  {N{\"u}sse}, \citenamefont {Schoengen}, \citenamefont {L{\"o}chel},
  \citenamefont {Hanke}, \citenamefont {Bratschitsch}, \citenamefont
  {Leitenstorfer}, \citenamefont {Aichele},\ and\ \citenamefont
  {Benson}}]{Wolters2012}%
  \BibitemOpen
  \bibfield  {author} {\bibinfo {author} {\bibfnamefont {J.}~\bibnamefont
  {Wolters}}, \bibinfo {author} {\bibfnamefont {G.}~\bibnamefont {Kewes}},
  \bibinfo {author} {\bibfnamefont {A.~W.}\ \bibnamefont {Schell}}, \bibinfo
  {author} {\bibfnamefont {N.}~\bibnamefont {N{\"u}sse}}, \bibinfo {author}
  {\bibfnamefont {M.}~\bibnamefont {Schoengen}}, \bibinfo {author}
  {\bibfnamefont {B.}~\bibnamefont {L{\"o}chel}}, \bibinfo {author}
  {\bibfnamefont {T.}~\bibnamefont {Hanke}}, \bibinfo {author} {\bibfnamefont
  {R.}~\bibnamefont {Bratschitsch}}, \bibinfo {author} {\bibfnamefont
  {A.}~\bibnamefont {Leitenstorfer}}, \bibinfo {author} {\bibfnamefont
  {T.}~\bibnamefont {Aichele}}, \ and\ \bibinfo {author} {\bibfnamefont
  {O.}~\bibnamefont {Benson}},\ }\href {\doibase 10.1002/pssb.201100156}
  {\bibfield  {journal} {\bibinfo  {journal} {physica status solidi (b)}\
  }\textbf {\bibinfo {volume} {249}},\ \bibinfo {pages} {918} (\bibinfo {year}
  {2012})}\BibitemShut {NoStop}%
\bibitem [{\citenamefont {Babinec}\ \emph {et~al.}(2010)\citenamefont
  {Babinec}, \citenamefont {Hausmann}, \citenamefont {Khan}, \citenamefont
  {Zhang}, \citenamefont {Maze}, \citenamefont {Hemmer},\ and\ \citenamefont
  {Loncar}}]{Babinec2010}%
  \BibitemOpen
  \bibfield  {author} {\bibinfo {author} {\bibfnamefont {T.~M.}\ \bibnamefont
  {Babinec}}, \bibinfo {author} {\bibfnamefont {B.~J.~M.}\ \bibnamefont
  {Hausmann}}, \bibinfo {author} {\bibfnamefont {M.}~\bibnamefont {Khan}},
  \bibinfo {author} {\bibfnamefont {Y.}~\bibnamefont {Zhang}}, \bibinfo
  {author} {\bibfnamefont {J.~R.}\ \bibnamefont {Maze}}, \bibinfo {author}
  {\bibfnamefont {P.~R.}\ \bibnamefont {Hemmer}}, \ and\ \bibinfo {author}
  {\bibfnamefont {M.}~\bibnamefont {Loncar}},\ }\href {\doibase
  10.1038/nnano.2010.6} {\bibfield  {journal} {\bibinfo  {journal} {Nat.
  Nanotechnol.}\ }\textbf {\bibinfo {volume} {5}},\ \bibinfo {pages} {195}
  (\bibinfo {year} {2010})}\BibitemShut {NoStop}%
\bibitem [{\citenamefont {Neu}\ \emph {et~al.}(2014)\citenamefont {Neu},
  \citenamefont {Appel}, \citenamefont {Ganzhorn}, \citenamefont
  {Miguel-S\'{a}nchez}, \citenamefont {Lesik}, \citenamefont {Mille},
  \citenamefont {Jacques}, \citenamefont {Tallaire}, \citenamefont {Achard},\
  and\ \citenamefont {Maletinsky}}]{Neu2014}%
  \BibitemOpen
  \bibfield  {author} {\bibinfo {author} {\bibfnamefont {E.}~\bibnamefont
  {Neu}}, \bibinfo {author} {\bibfnamefont {P.}~\bibnamefont {Appel}}, \bibinfo
  {author} {\bibfnamefont {M.}~\bibnamefont {Ganzhorn}}, \bibinfo {author}
  {\bibfnamefont {J.}~\bibnamefont {Miguel-S\'{a}nchez}}, \bibinfo {author}
  {\bibfnamefont {M.}~\bibnamefont {Lesik}}, \bibinfo {author} {\bibfnamefont
  {V.}~\bibnamefont {Mille}}, \bibinfo {author} {\bibfnamefont
  {V.}~\bibnamefont {Jacques}}, \bibinfo {author} {\bibfnamefont
  {A.}~\bibnamefont {Tallaire}}, \bibinfo {author} {\bibfnamefont
  {J.}~\bibnamefont {Achard}}, \ and\ \bibinfo {author} {\bibfnamefont
  {P.}~\bibnamefont {Maletinsky}},\ }\href {\doibase 10.1063/1.4871580}
  {\bibfield  {journal} {\bibinfo  {journal} {Appl. Phys. Lett.}\ }\textbf
  {\bibinfo {volume} {104}},\ \bibinfo {pages} {153108} (\bibinfo {year}
  {2014})}\BibitemShut {NoStop}%
\bibitem [{\citenamefont {Hausmann}\ \emph {et~al.}(2012)\citenamefont
  {Hausmann}, \citenamefont {Shields}, \citenamefont {Quan}, \citenamefont
  {Maletinsky}, \citenamefont {McCutcheon}, \citenamefont {Choy}, \citenamefont
  {Babinec}, \citenamefont {Kubanek}, \citenamefont {Yacoby}, \citenamefont
  {Lukin},\ and\ \citenamefont {Loncar}}]{Hausmann2012}%
  \BibitemOpen
  \bibfield  {author} {\bibinfo {author} {\bibfnamefont {B.~J.~M.}\
  \bibnamefont {Hausmann}}, \bibinfo {author} {\bibfnamefont {B.}~\bibnamefont
  {Shields}}, \bibinfo {author} {\bibfnamefont {Q.}~\bibnamefont {Quan}},
  \bibinfo {author} {\bibfnamefont {P.}~\bibnamefont {Maletinsky}}, \bibinfo
  {author} {\bibfnamefont {M.}~\bibnamefont {McCutcheon}}, \bibinfo {author}
  {\bibfnamefont {J.~T.}\ \bibnamefont {Choy}}, \bibinfo {author}
  {\bibfnamefont {T.~M.}\ \bibnamefont {Babinec}}, \bibinfo {author}
  {\bibfnamefont {A.}~\bibnamefont {Kubanek}}, \bibinfo {author} {\bibfnamefont
  {A.}~\bibnamefont {Yacoby}}, \bibinfo {author} {\bibfnamefont {M.~D.}\
  \bibnamefont {Lukin}}, \ and\ \bibinfo {author} {\bibfnamefont
  {M.}~\bibnamefont {Loncar}},\ }\href {\doibase 10.1021/nl204449n} {\bibfield
  {journal} {\bibinfo  {journal} {Nano Lett.}\ }\textbf {\bibinfo {volume}
  {12}},\ \bibinfo {pages} {1578} (\bibinfo {year} {2012})}\BibitemShut
  {NoStop}%
\bibitem [{\citenamefont {Faraon}\ \emph {et~al.}(2011)\citenamefont {Faraon},
  \citenamefont {Barclay}, \citenamefont {Santori}, \citenamefont {Fu},\ and\
  \citenamefont {Beausoleil}}]{Faraon2011}%
  \BibitemOpen
  \bibfield  {author} {\bibinfo {author} {\bibfnamefont {A.}~\bibnamefont
  {Faraon}}, \bibinfo {author} {\bibfnamefont {P.~E.}\ \bibnamefont {Barclay}},
  \bibinfo {author} {\bibfnamefont {C.}~\bibnamefont {Santori}}, \bibinfo
  {author} {\bibfnamefont {K.-M.~C.}\ \bibnamefont {Fu}}, \ and\ \bibinfo
  {author} {\bibfnamefont {R.~G.}\ \bibnamefont {Beausoleil}},\ }\href
  {\doibase 10.1038/nphoton.2011.52} {\bibfield  {journal} {\bibinfo  {journal}
  {Nat. Photonics}\ }\textbf {\bibinfo {volume} {5}},\ \bibinfo {pages} {301}
  (\bibinfo {year} {2011})}\BibitemShut {NoStop}%
\bibitem [{\citenamefont {Choy}\ \emph {et~al.}(2011)\citenamefont {Choy},
  \citenamefont {Hausmann}, \citenamefont {Babinec}, \citenamefont {Bulu},
  \citenamefont {Khan}, \citenamefont {Maletinsky}, \citenamefont {Yacoby},\
  and\ \citenamefont {Lon\v{c}ar}}]{Choy2011}%
  \BibitemOpen
  \bibfield  {author} {\bibinfo {author} {\bibfnamefont {J.~T.}\ \bibnamefont
  {Choy}}, \bibinfo {author} {\bibfnamefont {B.~J.~M.}\ \bibnamefont
  {Hausmann}}, \bibinfo {author} {\bibfnamefont {T.~M.}\ \bibnamefont
  {Babinec}}, \bibinfo {author} {\bibfnamefont {I.}~\bibnamefont {Bulu}},
  \bibinfo {author} {\bibfnamefont {M.}~\bibnamefont {Khan}}, \bibinfo {author}
  {\bibfnamefont {P.}~\bibnamefont {Maletinsky}}, \bibinfo {author}
  {\bibfnamefont {A.}~\bibnamefont {Yacoby}}, \ and\ \bibinfo {author}
  {\bibfnamefont {M.}~\bibnamefont {Lon\v{c}ar}},\ }\href {\doibase
  10.1038/nphoton.2011.249} {\bibfield  {journal} {\bibinfo  {journal} {Nat.
  Photonics}\ }\textbf {\bibinfo {volume} {5}},\ \bibinfo {pages} {738}
  (\bibinfo {year} {2011})}\BibitemShut {NoStop}%
\bibitem [{\citenamefont {Lee}\ \emph {et~al.}(2011)\citenamefont {Lee},
  \citenamefont {Chen}, \citenamefont {Eghlidi}, \citenamefont {Kukura},
  \citenamefont {Lettow}, \citenamefont {Renn}, \citenamefont {Sandoghdar},\
  and\ \citenamefont {G\"{o}tzinger}}]{Lee2011}%
  \BibitemOpen
  \bibfield  {author} {\bibinfo {author} {\bibfnamefont {K.~G.}\ \bibnamefont
  {Lee}}, \bibinfo {author} {\bibfnamefont {X.~W.}\ \bibnamefont {Chen}},
  \bibinfo {author} {\bibfnamefont {H.}~\bibnamefont {Eghlidi}}, \bibinfo
  {author} {\bibfnamefont {P.}~\bibnamefont {Kukura}}, \bibinfo {author}
  {\bibfnamefont {R.}~\bibnamefont {Lettow}}, \bibinfo {author} {\bibfnamefont
  {A.}~\bibnamefont {Renn}}, \bibinfo {author} {\bibfnamefont {V.}~\bibnamefont
  {Sandoghdar}}, \ and\ \bibinfo {author} {\bibfnamefont {S.}~\bibnamefont
  {G\"{o}tzinger}},\ }\href {\doibase 10.1038/nphoton.2010.312} {\bibfield
  {journal} {\bibinfo  {journal} {Nat. Photonics}\ }\textbf {\bibinfo {volume}
  {5}},\ \bibinfo {pages} {166} (\bibinfo {year} {2011})}\BibitemShut {NoStop}%
\bibitem [{\citenamefont {Chu}\ \emph {et~al.}(2014)\citenamefont {Chu},
  \citenamefont {Brenner}, \citenamefont {Chen}, \citenamefont {Ghosh},
  \citenamefont {Hollingsworth}, \citenamefont {Sandoghdar},\ and\
  \citenamefont {Goetzinger}}]{Chu2014a}%
  \BibitemOpen
  \bibfield  {author} {\bibinfo {author} {\bibfnamefont {X.~L.}\ \bibnamefont
  {Chu}}, \bibinfo {author} {\bibfnamefont {T.~J.~K.}\ \bibnamefont {Brenner}},
  \bibinfo {author} {\bibfnamefont {X.~W.}\ \bibnamefont {Chen}}, \bibinfo
  {author} {\bibfnamefont {Y.}~\bibnamefont {Ghosh}}, \bibinfo {author}
  {\bibfnamefont {J.~A.}\ \bibnamefont {Hollingsworth}}, \bibinfo {author}
  {\bibfnamefont {V.}~\bibnamefont {Sandoghdar}}, \ and\ \bibinfo {author}
  {\bibfnamefont {S.}~\bibnamefont {Goetzinger}},\ }\href
  {http://arxiv.org/abs/1406.0626} {\  (\bibinfo {year} {2014})},\ \Eprint
  {http://arxiv.org/abs/1406.0626} {arXiv:1406.0626} \BibitemShut {NoStop}%
\bibitem [{\citenamefont {Luan}\ \emph {et~al.}(2006)\citenamefont {Luan},
  \citenamefont {Sievert}, \citenamefont {Watkins}, \citenamefont {Mu},
  \citenamefont {Hong},\ and\ \citenamefont {Ketterson}}]{Luan2006}%
  \BibitemOpen
  \bibfield  {author} {\bibinfo {author} {\bibfnamefont {L.}~\bibnamefont
  {Luan}}, \bibinfo {author} {\bibfnamefont {P.~R.}\ \bibnamefont {Sievert}},
  \bibinfo {author} {\bibfnamefont {B.}~\bibnamefont {Watkins}}, \bibinfo
  {author} {\bibfnamefont {W.}~\bibnamefont {Mu}}, \bibinfo {author}
  {\bibfnamefont {Z.}~\bibnamefont {Hong}}, \ and\ \bibinfo {author}
  {\bibfnamefont {J.~B.}\ \bibnamefont {Ketterson}},\ }\href {\doibase
  10.1063/1.2234299} {\bibfield  {journal} {\bibinfo  {journal} {Appl. Phys.
  Lett.}\ }\textbf {\bibinfo {volume} {89}},\ \bibinfo {pages} {031119}
  (\bibinfo {year} {2006})}\BibitemShut {NoStop}%
\bibitem [{\citenamefont {Fu}\ \emph {et~al.}(2008)\citenamefont {Fu},
  \citenamefont {Santori}, \citenamefont {Barclay}, \citenamefont
  {Aharonovich}, \citenamefont {Prawer}, \citenamefont {Meyer}, \citenamefont
  {Holm},\ and\ \citenamefont {Beausoleil}}]{Fu2008}%
  \BibitemOpen
  \bibfield  {author} {\bibinfo {author} {\bibfnamefont {K.-M.~C.}\
  \bibnamefont {Fu}}, \bibinfo {author} {\bibfnamefont {C.}~\bibnamefont
  {Santori}}, \bibinfo {author} {\bibfnamefont {P.~E.}\ \bibnamefont
  {Barclay}}, \bibinfo {author} {\bibfnamefont {I.}~\bibnamefont
  {Aharonovich}}, \bibinfo {author} {\bibfnamefont {S.}~\bibnamefont {Prawer}},
  \bibinfo {author} {\bibfnamefont {N.}~\bibnamefont {Meyer}}, \bibinfo
  {author} {\bibfnamefont {A.~M.}\ \bibnamefont {Holm}}, \ and\ \bibinfo
  {author} {\bibfnamefont {R.~G.}\ \bibnamefont {Beausoleil}},\ }\href
  {\doibase 10.1063/1.3045950} {\bibfield  {journal} {\bibinfo  {journal}
  {Appl. Phys. Lett.}\ }\textbf {\bibinfo {volume} {93}},\ \bibinfo {pages}
  {234107} (\bibinfo {year} {2008})}\BibitemShut {NoStop}%
\bibitem [{\citenamefont {Wu}\ \emph {et~al.}(1999)\citenamefont {Wu},
  \citenamefont {Feke}, \citenamefont {Grober},\ and\ \citenamefont
  {Ghislain}}]{Wu1999}%
  \BibitemOpen
  \bibfield  {author} {\bibinfo {author} {\bibfnamefont {Q.}~\bibnamefont
  {Wu}}, \bibinfo {author} {\bibfnamefont {G.~D.}\ \bibnamefont {Feke}},
  \bibinfo {author} {\bibfnamefont {R.~D.}\ \bibnamefont {Grober}}, \ and\
  \bibinfo {author} {\bibfnamefont {L.~P.}\ \bibnamefont {Ghislain}},\ }\href
  {\doibase 10.1063/1.125537} {\bibfield  {journal} {\bibinfo  {journal} {Appl.
  Phys. Lett.}\ }\textbf {\bibinfo {volume} {75}},\ \bibinfo {pages} {4064}
  (\bibinfo {year} {1999})}\BibitemShut {NoStop}%
\bibitem [{\citenamefont {Suppl.}()}]{Suppl}%
  \BibitemOpen
  \bibfield  {author} {\bibinfo {author} {\bibnamefont {See supplementary information}}}\href@noop
  {} {\ }\BibitemShut {NoStop}%
\bibitem [{\citenamefont {Courtois}\ \emph {et~al.}(1996)\citenamefont
  {Courtois}, \citenamefont {Courty},\ and\ \citenamefont
  {Mertz}}]{Courtois1996}%
  \BibitemOpen
  \bibfield  {author} {\bibinfo {author} {\bibfnamefont {J.}~\bibnamefont
  {Courtois}}, \bibinfo {author} {\bibfnamefont {J.}~\bibnamefont {Courty}}, \
  and\ \bibinfo {author} {\bibfnamefont {J.}~\bibnamefont {Mertz}},\ }\href
  {http://www.ncbi.nlm.nih.gov/pubmed/9913082} {\bibfield  {journal} {\bibinfo
  {journal} {Phys. Rev. A}\ }\textbf {\bibinfo {volume} {53}},\ \bibinfo
  {pages} {1862} (\bibinfo {year} {1996})}\BibitemShut {NoStop}%
\bibitem [{\citenamefont {Lieb}\ \emph {et~al.}(2004)\citenamefont {Lieb},
  \citenamefont {Zavislan},\ and\ \citenamefont {Novotny}}]{Lieb2004}%
  \BibitemOpen
  \bibfield  {author} {\bibinfo {author} {\bibfnamefont {M.~A.}\ \bibnamefont
  {Lieb}}, \bibinfo {author} {\bibfnamefont {J.~M.}\ \bibnamefont {Zavislan}},
  \ and\ \bibinfo {author} {\bibfnamefont {L.}~\bibnamefont {Novotny}},\ }\href
  {\doibase 10.1364/JOSAB.21.001210} {\bibfield  {journal} {\bibinfo  {journal}
  {J. Opt. Soc. Am. B}\ }\textbf {\bibinfo {volume} {21}},\ \bibinfo {pages}
  {1210} (\bibinfo {year} {2004})}\BibitemShut {NoStop}%
\bibitem [{\citenamefont {Bass}\ \emph {et~al.}(2009)\citenamefont {Bass},
  \citenamefont {DeCusatis}, \citenamefont {Enoch}, \citenamefont
  {Lakshminarayanan}, \citenamefont {Li}, \citenamefont {MacDonald},
  \citenamefont {Mahajan},\ and\ \citenamefont {Stryland}}]{Bass2009}%
  \BibitemOpen
  \bibfield  {author} {\bibinfo {author} {\bibfnamefont {M.}~\bibnamefont
  {Bass}}, \bibinfo {author} {\bibfnamefont {C.}~\bibnamefont {DeCusatis}},
  \bibinfo {author} {\bibfnamefont {J.~M.}\ \bibnamefont {Enoch}}, \bibinfo
  {author} {\bibfnamefont {V.}~\bibnamefont {Lakshminarayanan}}, \bibinfo
  {author} {\bibfnamefont {G.}~\bibnamefont {Li}}, \bibinfo {author}
  {\bibfnamefont {C.}~\bibnamefont {MacDonald}}, \bibinfo {author}
  {\bibfnamefont {V.~N.}\ \bibnamefont {Mahajan}}, \ and\ \bibinfo {author}
  {\bibfnamefont {E.~V.}\ \bibnamefont {Stryland}},\ }\href
  {http://books.google.com/books?id=1CES2tOBwikC\&pgis=1} {\emph {\bibinfo
  {title} {{Handbook of Optics, Third Edition Volume IV}}}}\ (\bibinfo {year}
  {2009})\ p.\ \bibinfo {pages} {1152}\BibitemShut {NoStop}%
\bibitem [{\citenamefont {Le~Sage}\ \emph {et~al.}(2013)\citenamefont
  {Le~Sage}, \citenamefont {Arai}, \citenamefont {Glenn}, \citenamefont
  {{DeVience}}, \citenamefont {Pham}, \citenamefont {{Rahn-Lee}}, \citenamefont
  {Lukin}, \citenamefont {Yacoby}, \citenamefont {Komeili},\ and\ \citenamefont
  {Walsworth}}]{Lesage2013}%
  \BibitemOpen
  \bibfield  {author} {\bibinfo {author} {\bibfnamefont {D.}~\bibnamefont
  {Le~Sage}}, \bibinfo {author} {\bibfnamefont {K.}~\bibnamefont {Arai}},
  \bibinfo {author} {\bibfnamefont {D.~R.}\ \bibnamefont {Glenn}}, \bibinfo
  {author} {\bibfnamefont {S.~J.}\ \bibnamefont {{DeVience}}}, \bibinfo
  {author} {\bibfnamefont {L.~M.}\ \bibnamefont {Pham}}, \bibinfo {author}
  {\bibfnamefont {L.}~\bibnamefont {{Rahn-Lee}}}, \bibinfo {author}
  {\bibfnamefont {M.~D.}\ \bibnamefont {Lukin}}, \bibinfo {author}
  {\bibfnamefont {A.}~\bibnamefont {Yacoby}}, \bibinfo {author} {\bibfnamefont
  {A.}~\bibnamefont {Komeili}}, \ and\ \bibinfo {author} {\bibfnamefont
  {R.~L.}\ \bibnamefont {Walsworth}},\ }\href {\doibase 10.1038/nature12072}
  {\bibfield  {journal} {\bibinfo  {journal} {Nature}\ }\textbf {\bibinfo
  {volume} {496}},\ \bibinfo {pages} {486} (\bibinfo {year}
  {2013})}\BibitemShut {NoStop}%
\bibitem [{\citenamefont {Staudacher}\ \emph {et~al.}(2013)\citenamefont
  {Staudacher}, \citenamefont {Shi}, \citenamefont {Pezzagna}, \citenamefont
  {Meijer}, \citenamefont {Du}, \citenamefont {Meriles}, \citenamefont
  {Reinhard},\ and\ \citenamefont {Wrachtrup}}]{Staudacher2013}%
  \BibitemOpen
  \bibfield  {author} {\bibinfo {author} {\bibfnamefont {T.}~\bibnamefont
  {Staudacher}}, \bibinfo {author} {\bibfnamefont {F.}~\bibnamefont {Shi}},
  \bibinfo {author} {\bibfnamefont {S.}~\bibnamefont {Pezzagna}}, \bibinfo
  {author} {\bibfnamefont {J.}~\bibnamefont {Meijer}}, \bibinfo {author}
  {\bibfnamefont {J.}~\bibnamefont {Du}}, \bibinfo {author} {\bibfnamefont
  {C.~A.}\ \bibnamefont {Meriles}}, \bibinfo {author} {\bibfnamefont
  {F.}~\bibnamefont {Reinhard}}, \ and\ \bibinfo {author} {\bibfnamefont
  {J.}~\bibnamefont {Wrachtrup}},\ }\href {\doibase 10.1126/science.1231675}
  {\bibfield  {journal} {\bibinfo  {journal} {Science}\ }\textbf {\bibinfo
  {volume} {339}},\ \bibinfo {pages} {561} (\bibinfo {year}
  {2013})}\BibitemShut {NoStop}%
\bibitem [{\citenamefont {Neu}\ \emph {et~al.}(2011)\citenamefont {Neu},
  \citenamefont {Steinmetz}, \citenamefont {Riedrich-M\"{o}ller}, \citenamefont
  {Gsell}, \citenamefont {Fischer}, \citenamefont {Schreck},\ and\
  \citenamefont {Becher}}]{Neu2011}%
  \BibitemOpen
  \bibfield  {author} {\bibinfo {author} {\bibfnamefont {E.}~\bibnamefont
  {Neu}}, \bibinfo {author} {\bibfnamefont {D.}~\bibnamefont {Steinmetz}},
  \bibinfo {author} {\bibfnamefont {J.}~\bibnamefont {Riedrich-M\"{o}ller}},
  \bibinfo {author} {\bibfnamefont {S.}~\bibnamefont {Gsell}}, \bibinfo
  {author} {\bibfnamefont {M.}~\bibnamefont {Fischer}}, \bibinfo {author}
  {\bibfnamefont {M.}~\bibnamefont {Schreck}}, \ and\ \bibinfo {author}
  {\bibfnamefont {C.}~\bibnamefont {Becher}},\ }\href {\doibase
  10.1088/1367-2630/13/2/025012} {\bibfield  {journal} {\bibinfo  {journal}
  {New J. Phys.}\ }\textbf {\bibinfo {volume} {13}},\ \bibinfo {pages} {025012}
  (\bibinfo {year} {2011})}\BibitemShut {NoStop}%
\bibitem [{\citenamefont {Riedel}\ \emph {et~al.}(2012)\citenamefont {Riedel},
  \citenamefont {Fuchs}, \citenamefont {Kraus}, \citenamefont {V\"{a}th},
  \citenamefont {Sperlich}, \citenamefont {Dyakonov}, \citenamefont
  {Soltamova}, \citenamefont {Baranov}, \citenamefont {Ilyin},\ and\
  \citenamefont {Astakhov}}]{Riedel2012a}%
  \BibitemOpen
  \bibfield  {author} {\bibinfo {author} {\bibfnamefont {D.}~\bibnamefont
  {Riedel}}, \bibinfo {author} {\bibfnamefont {F.}~\bibnamefont {Fuchs}},
  \bibinfo {author} {\bibfnamefont {H.}~\bibnamefont {Kraus}}, \bibinfo
  {author} {\bibfnamefont {S.}~\bibnamefont {V\"{a}th}}, \bibinfo {author}
  {\bibfnamefont {A.}~\bibnamefont {Sperlich}}, \bibinfo {author}
  {\bibfnamefont {V.}~\bibnamefont {Dyakonov}}, \bibinfo {author}
  {\bibfnamefont {A.~A.}\ \bibnamefont {Soltamova}}, \bibinfo {author}
  {\bibfnamefont {P.~G.}\ \bibnamefont {Baranov}}, \bibinfo {author}
  {\bibfnamefont {V.~A.}\ \bibnamefont {Ilyin}}, \ and\ \bibinfo {author}
  {\bibfnamefont {G.~V.}\ \bibnamefont {Astakhov}},\ }\href {\doibase
  10.1103/PhysRevLett.109.226402} {\bibfield  {journal} {\bibinfo  {journal}
  {Phys. Rev. Lett.}\ }\textbf {\bibinfo {volume} {109}},\ \bibinfo {pages}
  {226402} (\bibinfo {year} {2012})}\BibitemShut {NoStop}%
\bibitem [{\citenamefont {Alegre}\ \emph {et~al.}(2007)\citenamefont {Alegre},
  \citenamefont {Santori}, \citenamefont {Medeiros-Ribeiro},\ and\
  \citenamefont {Beausoleil}}]{Alegre2007}%
  \BibitemOpen
  \bibfield  {author} {\bibinfo {author} {\bibfnamefont {T.~P.~M.}\
  \bibnamefont {Alegre}}, \bibinfo {author} {\bibfnamefont {C.}~\bibnamefont
  {Santori}}, \bibinfo {author} {\bibfnamefont {G.}~\bibnamefont
  {Medeiros-Ribeiro}}, \ and\ \bibinfo {author} {\bibfnamefont {R.~G.}\
  \bibnamefont {Beausoleil}},\ }\href {\doibase 10.1103/PhysRevB.76.165205}
  {\bibfield  {journal} {\bibinfo  {journal} {Phys. Rev. B}\ }\textbf {\bibinfo
  {volume} {76}},\ \bibinfo {pages} {165205} (\bibinfo {year}
  {2007})}\BibitemShut {NoStop}%
\bibitem [{\citenamefont {Lee}\ \emph {et~al.}(2008)\citenamefont {Lee},
  \citenamefont {Gu}, \citenamefont {Dawson}, \citenamefont {Friel},\ and\
  \citenamefont {Scarsbrook}}]{Lee2008}%
  \BibitemOpen
  \bibfield  {author} {\bibinfo {author} {\bibfnamefont {C.}~\bibnamefont
  {Lee}}, \bibinfo {author} {\bibfnamefont {E.}~\bibnamefont {Gu}}, \bibinfo
  {author} {\bibfnamefont {M.}~\bibnamefont {Dawson}}, \bibinfo {author}
  {\bibfnamefont {I.}~\bibnamefont {Friel}}, \ and\ \bibinfo {author}
  {\bibfnamefont {G.}~\bibnamefont {Scarsbrook}},\ }\href {\doibase
  10.1016/j.diamond.2008.01.011} {\bibfield  {journal} {\bibinfo  {journal}
  {Diam. Relat. Mater.}\ }\textbf {\bibinfo {volume} {17}},\ \bibinfo {pages}
  {1292} (\bibinfo {year} {2008})}\BibitemShut {NoStop}%
\bibitem [{\citenamefont {Neu}\ \emph {et~al.}(2014)\citenamefont {Neu},
  \citenamefont {Appel}, \citenamefont {Ganzhorn}, \citenamefont
  {Miguel-S\'{a}nchez}, \citenamefont {Lesik}, \citenamefont {Mille},
  \citenamefont {Jacques}, \citenamefont {Tallaire}, \citenamefont {Achard},\
  and\ \citenamefont {Maletinsky}}]{Neu2014}%
  \BibitemOpen
  \bibfield  {author} {\bibinfo {author} {\bibfnamefont {E.}~\bibnamefont
  {Neu}}, \bibinfo {author} {\bibfnamefont {P.}~\bibnamefont {Appel}}, \bibinfo
  {author} {\bibfnamefont {M.}~\bibnamefont {Ganzhorn}}, \bibinfo {author}
  {\bibfnamefont {J.}~\bibnamefont {Miguel-S\'{a}nchez}}, \bibinfo {author}
  {\bibfnamefont {M.}~\bibnamefont {Lesik}}, \bibinfo {author} {\bibfnamefont
  {V.}~\bibnamefont {Mille}}, \bibinfo {author} {\bibfnamefont
  {V.}~\bibnamefont {Jacques}}, \bibinfo {author} {\bibfnamefont
  {A.}~\bibnamefont {Tallaire}}, \bibinfo {author} {\bibfnamefont
  {J.}~\bibnamefont {Achard}}, \ and\ \bibinfo {author} {\bibfnamefont
  {P.}~\bibnamefont {Maletinsky}},\ }\href {\doibase 10.1063/1.4871580}
  {\bibfield  {journal} {\bibinfo  {journal} {Appl. Phys. Lett.}\ }\textbf
  {\bibinfo {volume} {104}},\ \bibinfo {pages} {153108} (\bibinfo {year}
  {2014})}\BibitemShut {NoStop}%
\bibitem [{\citenamefont {{L Luan and P R Sievert and J B
  Ketterson}}(2006)}]{Luan2006b}%
  \BibitemOpen
  \bibfield  {author} {\bibinfo {author} {\bibnamefont {{L Luan and P R Sievert
  and J B Ketterson}}},\ }\href {\doibase
  http://dx.doi.org/10.1088/1367-2630/8/11/264} {\bibfield  {journal} {\bibinfo
   {journal} {New J. Phys.}\ }\textbf {\bibinfo {volume} {8}},\ \bibinfo
  {pages} {264} (\bibinfo {year} {2006})}\BibitemShut {NoStop}%
\bibitem [{\citenamefont {Lieb}\ \emph {et~al.}(2004)\citenamefont {Lieb},
  \citenamefont {Zavislan},\ and\ \citenamefont {Novotny}}]{Lieb2004}%
  \BibitemOpen
  \bibfield  {author} {\bibinfo {author} {\bibfnamefont {M.~A.}\ \bibnamefont
  {Lieb}}, \bibinfo {author} {\bibfnamefont {J.~M.}\ \bibnamefont {Zavislan}},
  \ and\ \bibinfo {author} {\bibfnamefont {L.}~\bibnamefont {Novotny}},\ }\href
  {\doibase 10.1364/JOSAB.21.001210} {\bibfield  {journal} {\bibinfo  {journal}
  {J. Opt. Soc. Am. B}\ }\textbf {\bibinfo {volume} {21}},\ \bibinfo {pages}
  {1210} (\bibinfo {year} {2004})}\BibitemShut {NoStop}%
\end{thebibliography}

\newpage

\section*{Supplementary Information}

Details of the fabrication process for the thin diamond membranes as well as their transfer to GaP solid immersion lenses (SILs) are presented. The calculations of the radiation pattern of the antenna are introduced. In addition, we present absorption measurements on the GaP material we employed.

\section*{S1. Sample fabrication}
\label{sec:fabrication}
\noindent\textbf{Sample details}\\
We use a single crystal, chemical vapor deposition diamond sample (thickness $\unit[21.4-21.9]{\mu m}$) double-side polished to an optical finish (typically $\mathrm{R_a<\unit[3]{nm}}$) from Delaware Diamond Knives, Inc. The sample consists of optical grade diamond with an impurity content of $\mathrm{[N_s]^0< \unit[1]{ppm}}$ and $\mathrm{[B]< \unit[5]{ppb}}$. Prior to processing, the sample is cleaned in boiling triacid (sulphuric acid, nitric acid and perchloric acid, mixed to 1:1:1) and rinsed in de-ionized water and solvents (acetone, ethanol, isopropanol) to remove any residual contamination from the polishing.\\

\begin{figure}
\includegraphics[width=8.6cm]{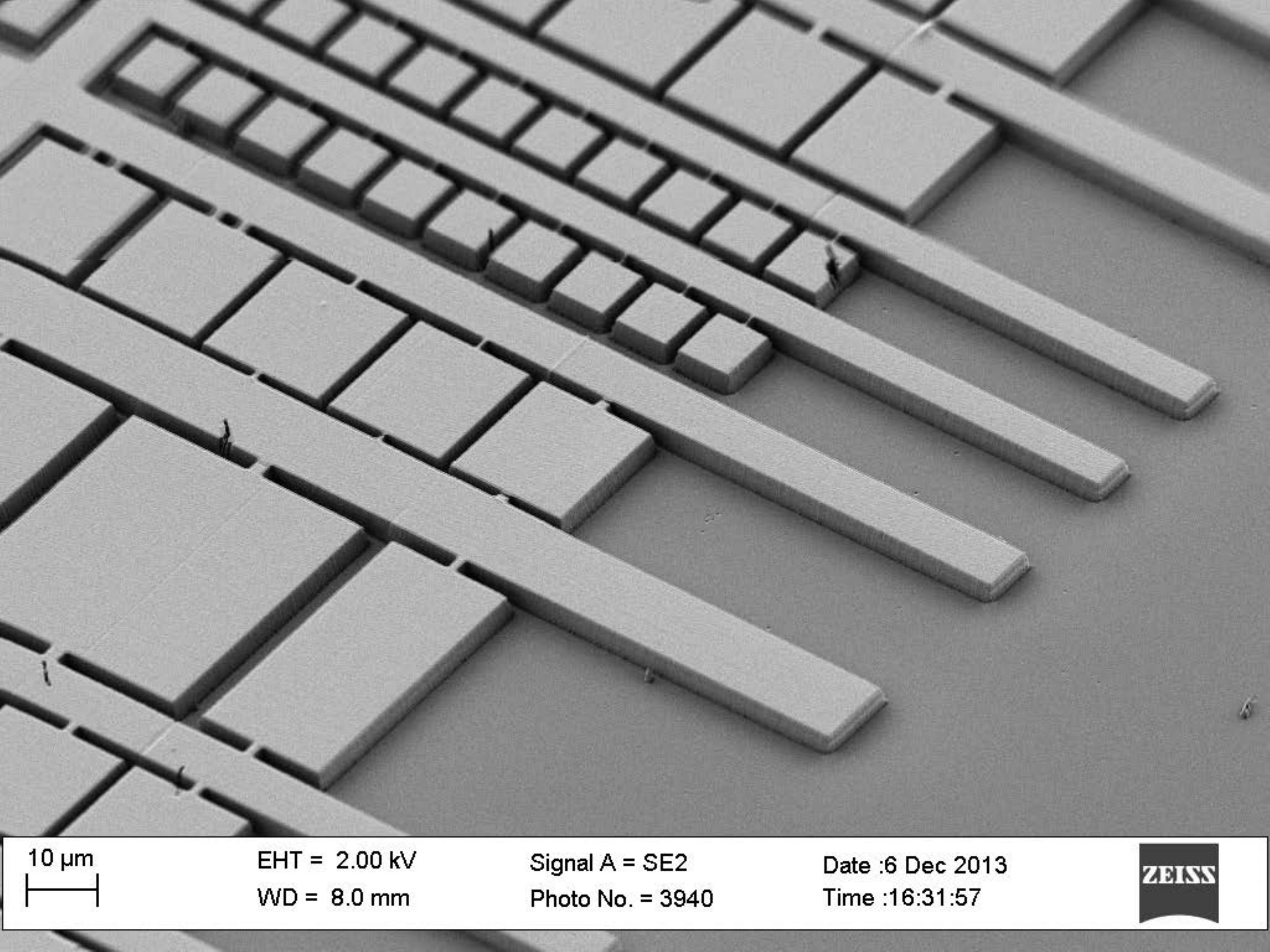}
\caption{SEM image of the FOX pattern transferred into the diamond. Note the membranes with different sizes as well as the holding structures.  \label{FOXpattern}}
\end{figure}
 
\begin{figure}
\includegraphics[width=8.6cm]{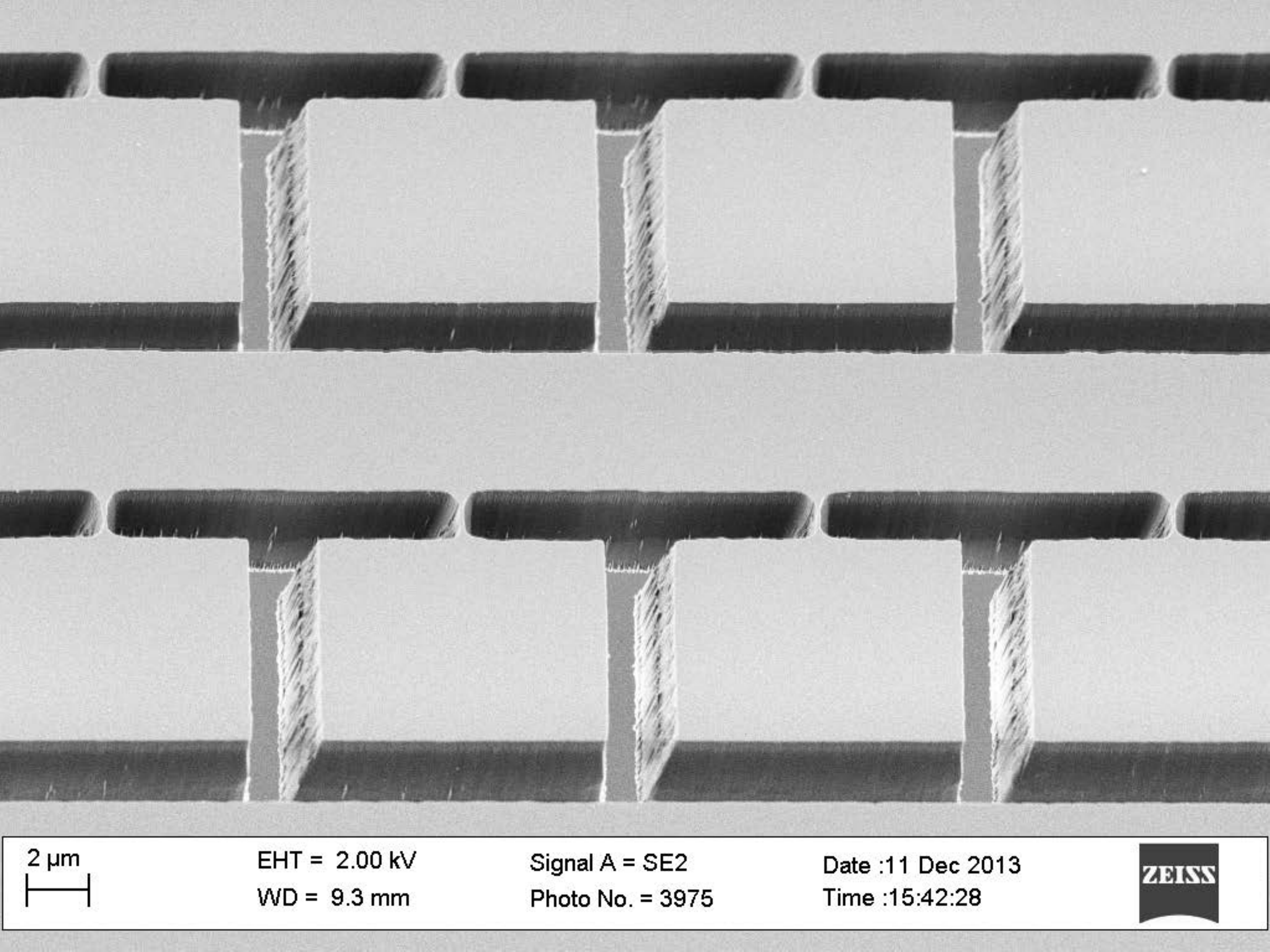}
\caption{SEM image of free standing membranes after mask removal and cleaning. \label{membranes}}
\end{figure}

\noindent\textbf{Fabrication of thin diamond layer}\\
To ease sample handling, the diamond is thinned down only in a small region of approximately $\unit[300\times600]{\mu m^2}$ using inductively coupled plasma reactive ion etching (ICP-RIE, Sentech SI 500). We employ an alternating plasma sequence with $\unit[10]{mins}$ of Ar/Cl$_2$ plasma (ICP power $\unit[400]{W}$, bias power $\unit[250]{W}$, reactor pressure $\unit[1]{Pa}$, $\unit[40]{sccm}$ Cl$_2$, $\unit[25]{sccm}$ Ar) followed by $\unit[20]{mins}$ of O$_2$ plasma (ICP power $\unit[700]{W}$, bias power $\unit[100]{W}$, reactor pressure $\unit[1.3]{Pa}$, $\unit[60]{sccm}$ O$_2$). The Ar/Cl$_2$ plasma aids in maintaining the surface quality throughout the long etching process \cite{Lee2008}. We estimate etch rates of $\unit[2.9]{\mu m/h}$ and $\unit[8.4]{\mu m/h}$ for the Ar/Cl$_2$ and the O$_2$ plasma, respectively. The diamond is etched down to a final thickness of $\unit[5.9-3.7]{\mu m}$ (local thickness variation inherent to etching/masking process). After etching, the sample is immersed in buffered oxide etch (BOE) solution and again cleaned as described in the previous paragraph. Scanning electron microscope (SEM) images reveal no resolvable roughness throughout the thinned region of the diamond.\\

\begin{figure}
\includegraphics[width=8.6cm]{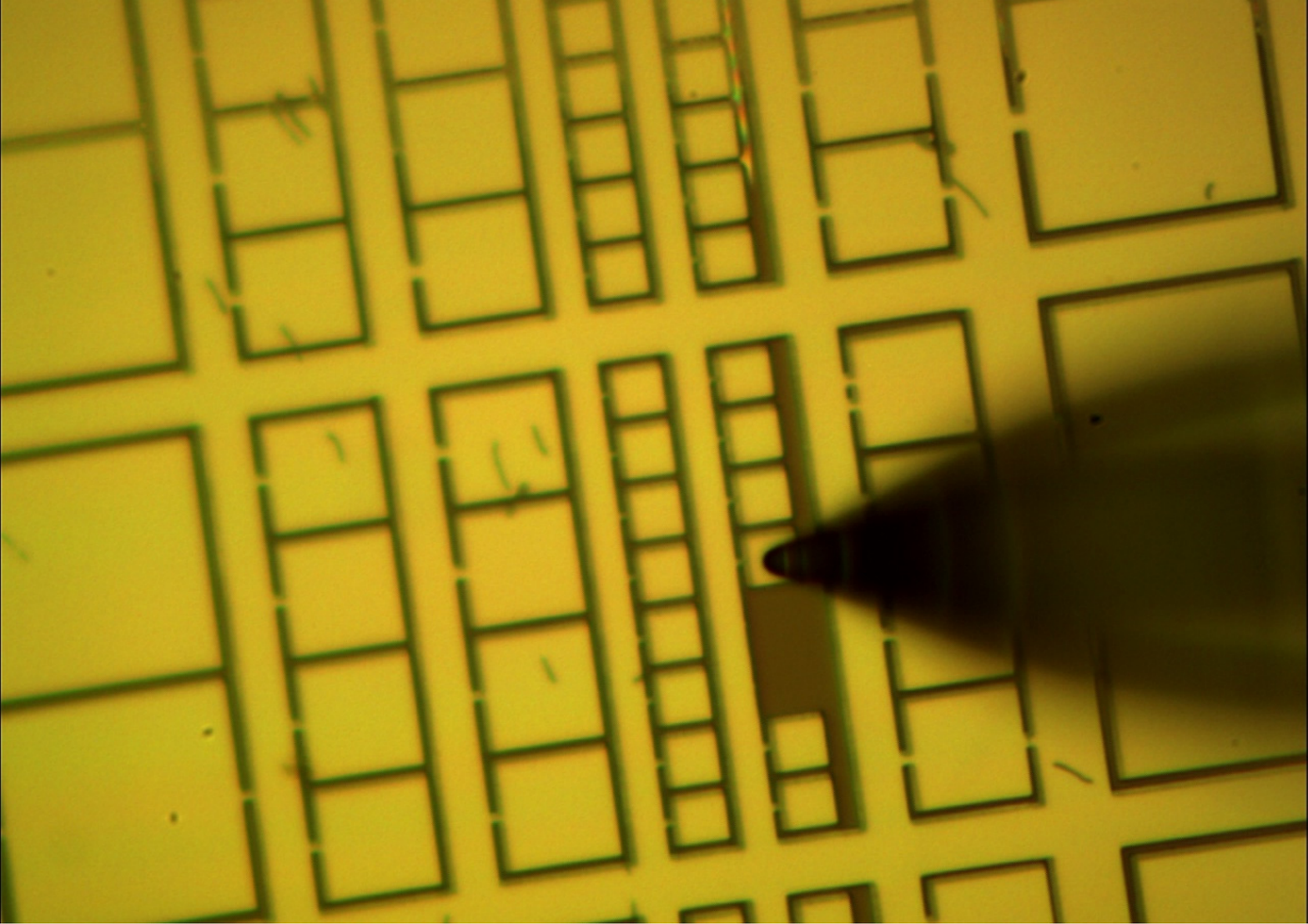}
\caption{Micromanipulator detaching of diamond membranes, the smallest with size around $\unit[10\times10]{\mu m^2}$ \label{micromanipulator}}
\end{figure}

\begin{figure}
\includegraphics[width=8.6cm]{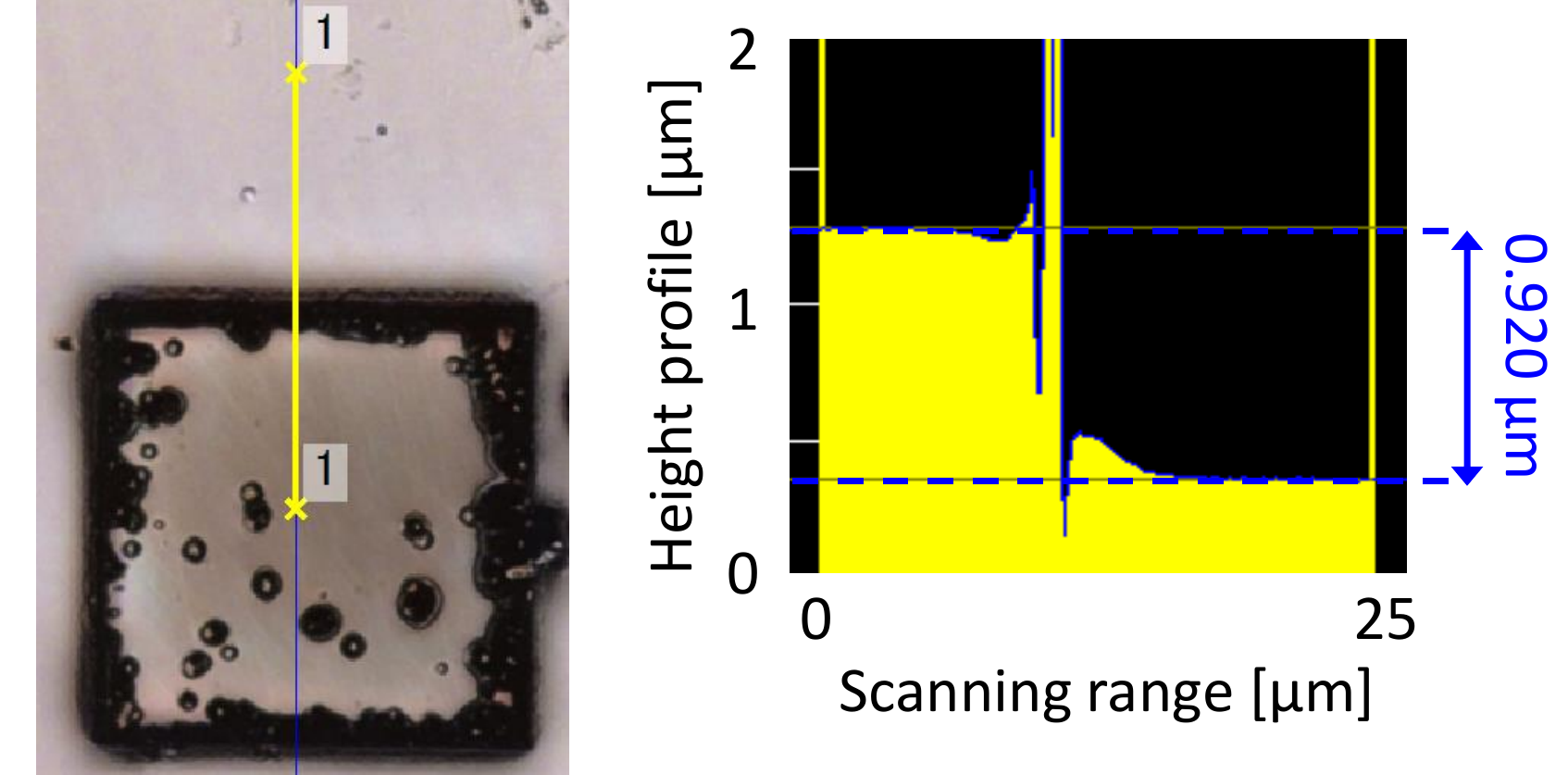}
\caption{Scanning laser confocal image of a thinned diamond membrane on GaP SIL. Note the blackened edges of the membrane and the occurrence of some needles. By measuring the step between the GaP and the diamond membrane's upper surface, we determine a thickness for the diamond membrane of $\approx \unit[0.920]{\mu m}$ \label{keyence_slabs}}
\end{figure}

\noindent\textbf{Fabrication of micro-structured diamond membranes for photonic device fabrication}\\
To ease placement and bonding of the diamond on the GaP SIL, the thin region of the diamond is structured to yield smaller diamond membranes (sizes between $\unit[10\times10]{\mu m^2}$ and $\unit[50\times50]{\mu m^2}$). To this end, we pattern the sample from the as yet non-etched side. We use electron beam lithography (30 keV) to pattern a layer of FOX-16 negative electron beam resist (Dow Corning) on a 2 nm Ti adhesion layer. For details of the mask fabrication see Ref.\ \cite{Neu2014}. To transfer the pattern into the diamond, we first use a $\unit[50]{s}$ Ar sputtering process in our ICP-RIE to remove the Ti adhesion layer (plasma parameters: 0.4 Pa, 50 sccm Ar, 500 W ICP, 300 W bias power). Diamond etching starts with a plasma containing 50\% Ar and O$_2$ respectively (gas flow $\unit[50]{sccm}$ each, $\unit[0.5]{Pa}$, ICP power $\unit[500]{W}$, bias power $\unit[200]{W}$). This plasma yields steep sidewalls and smooth surfaces in the non-masked areas \cite{Neu2014}. However, it also introduces a significant amount of FOX mask erosion. After $\unit[8]{mins}$ we thus change to a pure O$_2$ plasma (ICP power $\unit[700]{W}$, bias power $\unit[100]{W}$, reactor pressure $\unit[1.3]{Pa}$, $\unit[60]{sccm}$ O$_2$) to decrease mask erosion and etch through the diamond to yield free standing membranes. Part of the pattern can be seen in Fig.\,\ref{FOXpattern}. After the etching, the FOX mask is removed in BOE and the sample is acid cleaned again. Examples of free standing membranes are shown in Fig.\,\ref{membranes}.\\

\noindent\textbf{Transfer of membranes and thickness tuning}\\
The pre-defined membranes are detached from the sample using the glass tip of a micromanipulator (Narishige MMO-202ND), see Fig.\,\ref{micromanipulator}. They are placed at the center of the SIL using the micromanipulator tip and bond by van der Waals forces. To ensure optimal performance of the dielectric antenna devices and to investigate antenna performance in detail, the transferred membranes are thinned using ICP-RIE while attached to the SIL leading to thin membranes with well controlled thickness. This is feasible as the O$_2$ plasmas introduced above attack the GaP SIL only weakly (estimated etch rate pure O$_2$ plasma $\unit[2.5]{nm/min}$). To ensure the optimal thickness of the diamond, short etch steps are performed and the thickness of the fabricated diamond membrane is measured \textit{ex situ} using a laser-scanning confocal-microscope (Keyence, VK-X200K). Fig.\,\ref{keyence_slabs} shows an optical image of a diamond membrane on the SIL and the corresponding thickness measurement.

\begin{figure}
\includegraphics[width=8.6cm]{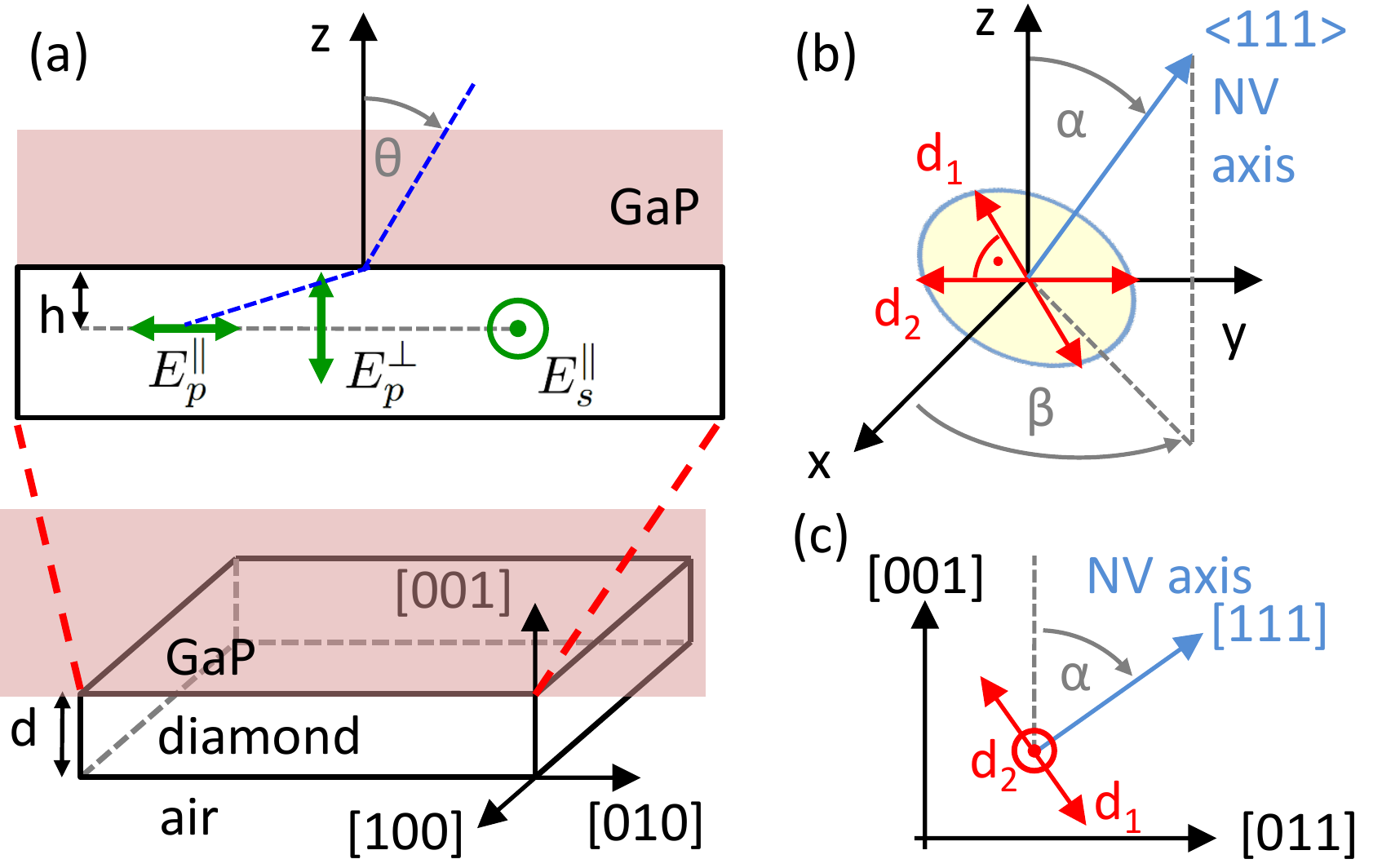}
\caption{\label{simulation} (a) Sketch of the dielectric antenna structure: a thin diamond membrane with thickness $d$ ($n \sim 2.4$) containing a single NV center is bonded to gallium phosphide (GaP, $n \sim 3.3$). The NV center is located at distance $h$ from the GaP-diamond interface. The dependence of the emission pattern from the polar angle $\theta$ can be described in terms of the three basic dipoles $E^p_\parallel(\theta)$, $E^p_\perp(\theta)$ and $E^s_\parallel(\theta)$. (b) Definition of the spherical coordinate system $(\alpha,\beta)$ used to describe the direction of the dipoles. (c) Particular orientation for the dipoles d$_1$ (in z-y plane, $\alpha=\arctan{1/\sqrt{2}}$ tilted from the z-axis) and d$_2$ (aligned along the y-axis).}
\end{figure}

\section*{S2. Calculations and estimation of collection efficiency}
\label{sec:calculation}
\noindent\textbf{Description of calculations}\\
The calculation of the emission pattern of our antenna structures relies on an asymptotic approach based on the Lorentz reciprocity theorem \cite{Courtois1996}. As a first step, we calculate the Fresnel coefficients for our 3 layer system consisting of a thin (100)-oriented diamond layer ($n\sim2.4$) located between a GaP ($n\sim3.3$) half-space and an air half-space ($n=1$), Fig.\,\ref{simulation}(a).

We determine the angular distribution of the emission of a dipole embedded in a diamond layer of thickness $d$ at distance $h$ from the GaP-diamond interface for a given wavelength $\lambda$. To this end, we calculate the electric field amplitude of a plane wave approaching the structure from infinity for different polarizations. We obtain the basic angular dependence of the forward $(+)$ and backward $(-)$ propagating s/p-polarized electric fields
$E^{(+)}_{s/p}(\theta)$ and $E^{(-)}_{s/p}(\theta)$
using the transfer matrix formalism, as proposed in Ref. \cite{Luan2006b}. Then, we derive the three basic quantities $E^p_\parallel(\theta)$, $E^s_\parallel(\theta)$ and $E^p_\perp(\theta)$, which describe the $\theta$-dependence of the electric field of the three possible dipole orientations, Fig.\,\ref{simulation}(a), for $\phi=0$:
\begin{equation}
\begin{aligned}
& E_s^\parallel(\theta)= E^{(+)}_{s}(\theta)+E^{(-)}_{s}(\theta)\\
& E_p^\parallel(\theta)=E^{(+)}_{p}(\theta)-E^{(-)}_{p}(\theta)\\
& E_p^\perp(\theta)=E^{(+)}_{p}(\theta)+E^{(-)}_{p}(\theta)
\end{aligned}
\end{equation}
While the radiation pattern is described by $(\theta,\phi)$, we introduce the spherical coordinates of the dipole orientation as $(\alpha,\beta)$ in order to describe the radiation pattern of a skewed dipole Fig.\,\ref{simulation}(b). The p- and s- polarized components of the electric field are then given by
\begin{equation}
\begin{aligned}
& E_p(\theta,\phi,\alpha,\beta)=E_p^\perp(\theta)\cos{\alpha}\sin{\theta},\\
& \hspace{2.5cm}       +E_p^\parallel(\theta)\sin{\alpha}\cos{\theta}\cos{(\phi-\beta)},\\
& E_s(\theta,\phi,\alpha,\beta)=E_s^\parallel(\theta)\sin{\alpha}\sin{(\phi-\beta).}
\end{aligned}
\end{equation}
The emitted power density per solid angle $d\Omega$ is then calculated via
\begin{equation}
P(\theta,\phi,\alpha,\beta) \propto n(\theta)\left(E_p E_p^* + E_s E_s^*\right)
\end{equation}
where $n(\theta)=n_{GaP}$ for $0<\theta,\pi/2$ and $n(\theta)=n_{air}$ for $\pi/2<\theta,\pi$. In order to calculate the emission pattern in the back focal plane, the standard apodization factor $\cos^{-1}{\theta}$ is introduced~\cite{Lieb2004}:
\begin{equation}
I(\theta,\phi,\alpha,\beta) \propto \frac{1}{\cos{\theta}}P(\theta,\phi,\alpha,\beta).
\end{equation}\\

\begin{figure}
\includegraphics[width=8.6cm]{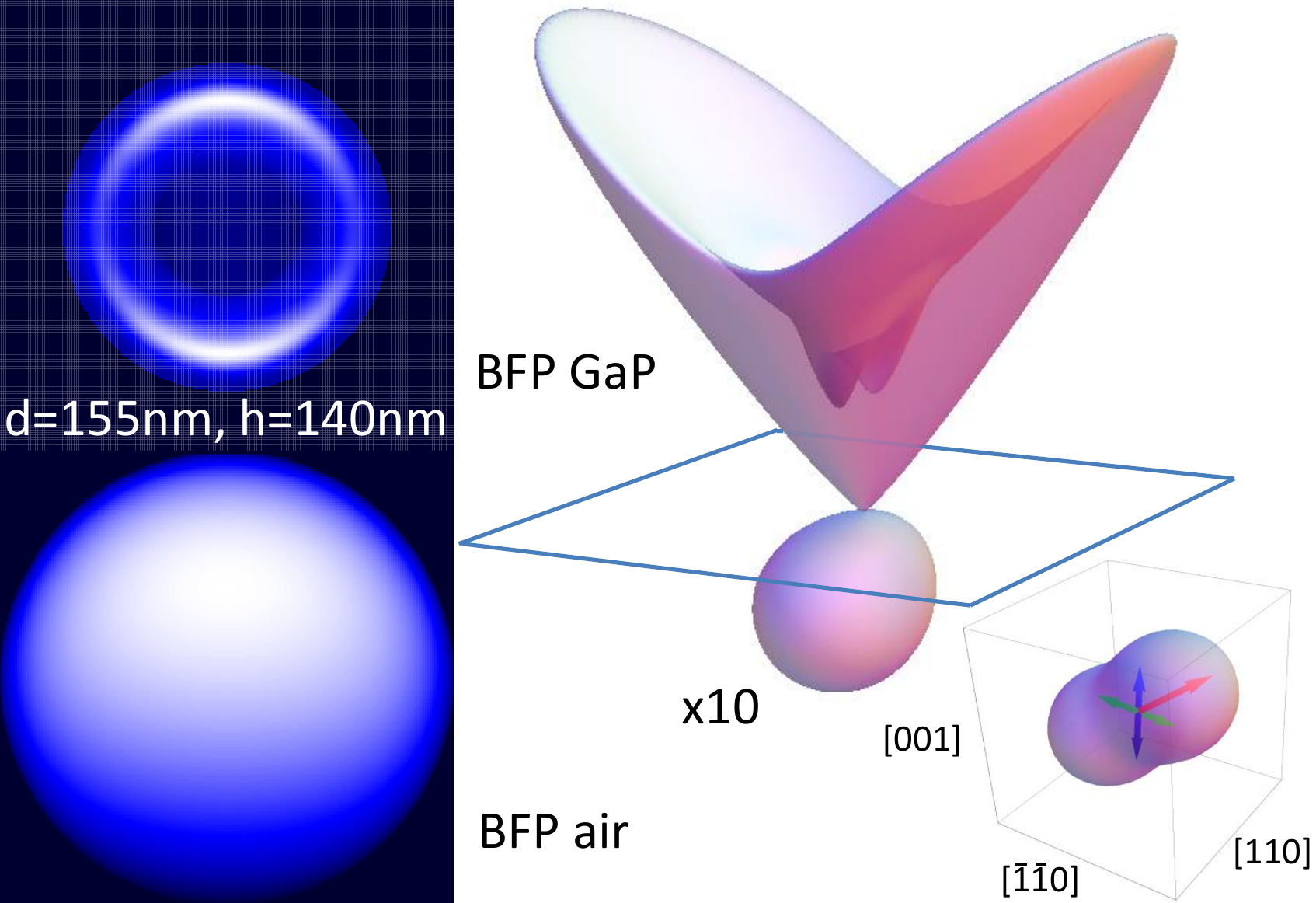}
\caption{\label{simulationNV} Simulated back focal plane images for an NV center with $d=155~$nm and $h=140~$ for emission into air and GaP. On the right, the corresponding emission pattern along with the emission pattern of the two NV dipoles in vacuum are shown.}
\end{figure}

\begin{figure*}
\includegraphics[width=14.33cm]{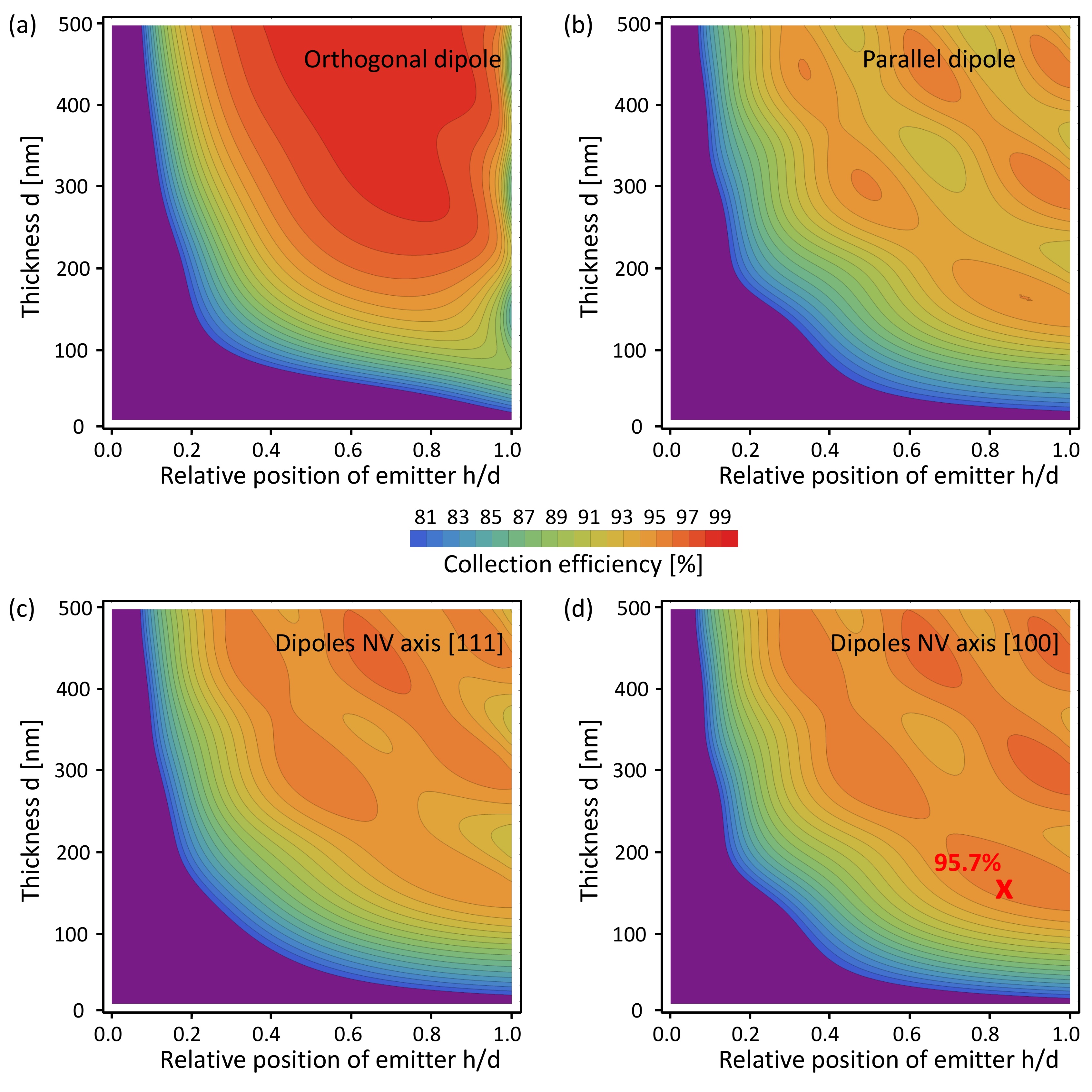}
\caption{\label{collection} Collection efficiency at $\lambda=675~$nm for different pairs of $h$ and $d$ for dipoles aligned (a) orthogonal, and (b) parallel to the GaP-diamond interface; and for NV emission in (c) [111] diamond, and (d) [100] diamond.}
\end{figure*}

\noindent\textbf{Calculation of the BFP for a single NV}\\
Fig.\,\ref{simulation}(c) displays the alignment of the optical dipoles in the plane normal to the NV axis. We note that, assuming an equal oscillator strength, the alignment of the dipoles within this plane can be chosen arbitrarily. We account for the NV PL spectrum by introducing a wavelength-dependent weighting factor for the emission profile and discretize the spectrum in 5~nm steps. We present the theoretical collection efficiencies  $P_{obj}/( P_{GaP}+ P_{air})$ for the two basic dipole orientations $P_\parallel$ and $P_\perp$ within the GaP-diamond dielectric antenna for different thicknesses $d$ and relative positions $h/d$. We calculate the total power emitted by integrating the emission profile over the relevant solid angle:
\begin{equation}
\begin{aligned}
& P_{obj}=\int^{2\pi}_{\phi=0}\int^{\arcsin{0.8}}_{\theta=0}P(\theta,\phi,\alpha,\beta)\sin{\theta}d\theta d\phi, \\
& P_{GaP}=\int^{2\pi}_{\phi=0}\int^{\pi/2}_{\theta=0}P(\theta,\phi,\alpha,\beta)\sin{\theta}d\theta d\phi, \\
& P_{air}=\int^{2\pi}_{\phi=0}\int^{\pi}_{\theta=\pi/2}P(\theta,\phi,\alpha,\beta)\sin{\theta}d\theta d\phi. \\
\end{aligned}
\end{equation}
Integrating both emission patterns over their respective half-space ($\phi=0...2\pi$, GaP: $\theta=0...\pi/2$, air: $\theta=\pi/2...\pi$) yields the total emitted power. Comparison with the power emitted up to the maximum the angle which can be collected by the objective ($\phi=0...2\pi$, $\theta=0...\arcsin(0.8)$) yields the collection efficiency. Fig.\,\ref{collection} displays the calculated collection efficiencies as a function of $d$ and $h/d$ for a dipole aligned orthogonal (a) and parallel (b) to the surface and for the NV emission in [111] diamond (c) and [100] diamond (d). The efficiencies are calculated for $\lambda=675~$nm. The theoretical collection efficiency is 95.7\% for an NV center in [100] diamond with $d=155~$nm and $h=140~$nm. Considering the entire broadband PL spectrum of an NV center, we calculate a collection efficiency of 95.1\% for the same parameters after weighting the calculation with the NV PL spectrum.

\begin{figure}
\includegraphics[width=8.6cm]{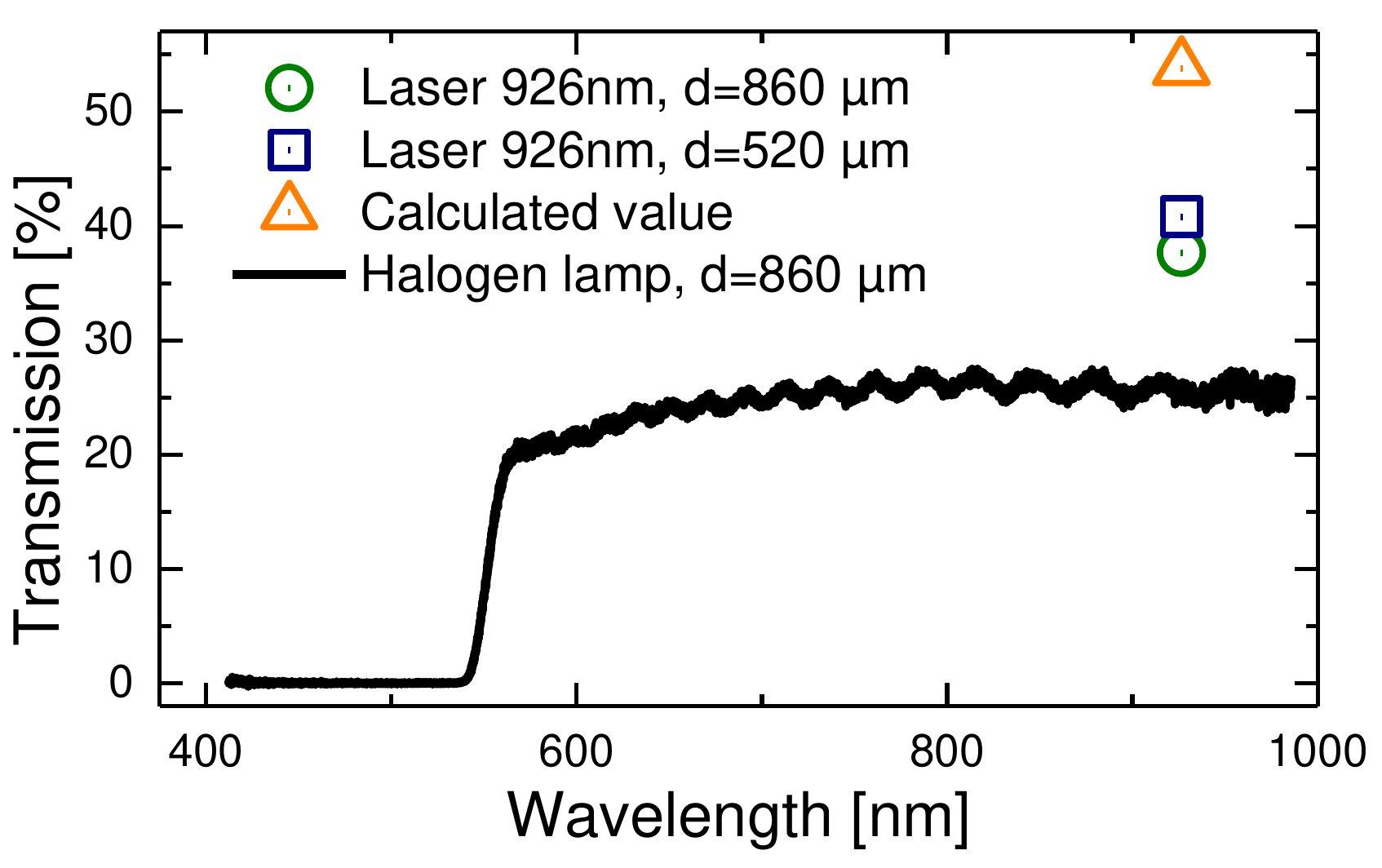}
\caption{\label{GaPabs} Transmission measurement of two GaP slabs (thickness d) performed using a halogen white-light source and a diode laser tuned to $\lambda=\unit[926]{nm}$. The calculated value gives the transmission estimated using Fresnel coefficients.}
\end{figure}

\section*{S3. Characterization of the absorption in the employed GaP material}
\label{sec:GaPabsorb}
The transmission of a $\unit[860]{\mu m}$ thick GaP slab was measured in order to estimate the absorption losses in the material (A.W.I.\ Industries). The solid line in Fig.\,\ref{GaPabs} shows the normalized transmission spectrum of white light passing through the GaP slab. The spectrum reveals a strong increase of transmission at $\unit[550]{nm}$ (associated with the GaP band gap) and a rather constant transmission in the region of interest from $\unit[600]{nm}$ to $\unit[750]{nm}$ and also at longer wavelengths. The oscillating behavior originates from interference effects.
Due to difficulties in collimating a white light source, this way of measurement implies a significant systematic error in the absolute transmission. Therefore, the transmission was also measured with coherent and collimated laser light, in this case at $\lambda=\unit[926]{nm}$. The resulting transmission of about 38\% is indicated in Fig.\,\ref{GaPabs} with a blue square. Without any scattering and without any absorption we would expect a transmission of 54\% at this wavelength (indicated with an orange triangle) simply from the reflection losses at the two GaP-air interfaces. The additional losses are related either to absorption/scattering in the material or to scattering at the surfaces. To differentiate between the two possibilities, we measured the transmission also through a thinner slab with $\unit[520]{\mu m}$ thickness. Figure\,\ref{GaPabs} shows that the losses reduce slightly, allowing us to deduce a loss coefficient of the GaP material, $\unit[0.098]{mm^{-1}}$, and a surface scattering loss of $\sim 4\%$ per GaP-air interface. We note that the SILs are polished slightly better than the slabs such that for the dielectric antenna structure, the main absorption/scattering loss arises from imperfections in the GaP material. 

\end{document}